\RequirePackage{fix-cm}
\documentclass[twocolumn,epjc3]{svjour3}  
\smartqed  
\usepackage{subcaption}
\RequirePackage{graphicx}
\RequirePackage{amsmath}
\usepackage{amssymb}
\usepackage{bigints}
\usepackage{adjustbox}
\usepackage{multirow}
\usepackage{array}
\usepackage{float}
\usepackage{cite}
\usepackage{balance}
\usepackage{rotating}

\usepackage[colorlinks=true,linkcolor=blue, citecolor=blue, urlcolor=blue]{hyperref}

\usepackage{orcidlink}
\usepackage{cite}
\usepackage{caption}
\usepackage{subcaption}
\usepackage{graphicx, amsmath}
\usepackage{float}
\usepackage{booktabs}
\usepackage{multirow}
\begin{document}

\title{Strong lensing and Hawking spectra of charged black hole under Lorentz violation theory
}


\author{Yenshembam Priyobarta Singh\orcidlink{0000-0003-3168-7493}\thanksref{e1,addr1} 
        \and Ningthoujam Media\orcidlink{0009-0007-7776-9511}\thanksref{e2,addr1}
        Telem Ibungochouba Singh\orcidlink{0000-0002-2568-0343}\thanksref{e3,addr1} 
}

\thankstext{e1}{e-mail: priyoyensh@gmail.com     }
\thankstext{e2}{e-mail: medyaningthoujam@gmail.com     }
\thankstext{e3}{e-mail: ibungochouba@rediffmail.com (corresponding author)}


\institute{Department of Mathematics, Manipur University, Canchipur 795003, India \label{addr1}
}


\maketitle

\begin{abstract}
In this paper, we investigate the strong gravitational lensing effects around the Reissner-Nordstr{\"o}m-like  black hole (RN-like BH)  in bumblebee gravity. We calculate the  lensing quantities such as deflection angle,  radius of the photon sphere, angular separation, relativistic image characteristics, Einstein ring and time delay. By taking the supermassive black holes (SMBHs) like  Sgr A*, M87*, NGC 1332, NGC 4649 etc. as RN-like BH, we compute the lensing observables and compare with the observables associated with the   Reissner-Nordström black hole (RNBH).
  The quasinormal modes (QNMs) of massless Dirac field perturbation of Reissner-Nordstr{\"o}m-de Sitter-like (RNdS-like) BH are also calculated. We discuss the behaviour of both real and imaginary parts of QNM frequencies  with varying of $Q$ and $L$. It shows  the damping rate and the oscillation frequency decrease with increasing $L$ but it has the opposite effect for the increase of $Q$.  We further explore the detectability of QNMs by LIGO and LISA, and study the impact of  $L$ and $Q$ on the detectable BH mass range. Further, we investigate the behaviour of greybody factor (GF) of RNdS-like BH in bumblebee gravity and find that the probability of wave transmission increases with increasing $L$ but decreases with the increase of $Q$. The behaviour of absorption cross section and sparsity of Hawking radiation for different values of $L$ and $Q$ are also analysed graphically.


\end{abstract}

\date{}
\section{Introduction}
A useful tool for determining the mass of dark gravitational bodies lying between the emitter and the observer is through a gravitational lensing. Gravitational lensing will occur when  a massive object like a BH, galaxy, or galaxy cluster lying between a light source and an observer  bends the light ray.
 The various observable effects such as magnification of the brightness of the source, multiple images of the same source and formation of ring structure around the lens can be studied from the gravitational bending of light rays.
Among the key predictions of general relativity (GR),  the deflection of light provided the first major experimental confirmation of the theory.
 Since the prediction of gravitational lensing within the framework of general relativity by Albert Einstein \cite{Einstein}, Ref. \cite{Walsh} observed and confirmed the gravitational lensing of the BH through astronomical observation. Depending on the degree of light bending, gravitational lensing can broadly be divided into weak and strong. When light passes far from the gravitational source, the weak deflection occurs but  when light travels close to the source it experiencing a significant deflection thereby causing strong lensing.
 In the weak field approximation, Schneider et al. \cite{Schneider} firstly studied the theory of gravitational lensing, providing a framework  to investigate the physical observations. Since then, many researchers developed various methods of weak gravitational lensing to study BH physics, focusing on aspects like the deflection angle and weak  lensing features of the RNBH by applying a perturbation method \cite{Sereno2004}, Taylor expansion formula for weak field limit of spherically symmetric BH \cite{Keeton} and modified to Kerr BH \cite{Sereno2006, Werner}, Gibbon-Werner method by applying Gauss-Bonnet theorem \cite{Bartelmann, Weinberg} and so on. \\
Strong gravitational lensing has been drawn a lot of interest in recent decades. It becomes a useful technique for studying the non luminous things like dark matter and extrasolar planets. 

The importance of studying the strong gravitational lensing is the possibility of exploring the spacetime geometry at the horizon of BH by applying the physical properties of the gravitational relativistic images. The strong gravitational lensing caused by compact objects  such as BHs and naked singularities, which posses photon spheres,  was initially investigated by Darwin \cite{Darwin}. Ref. \cite{Virbhadra} derived the lens equation after studying strong gravitational lensing. Ref. \cite{Frittelli} purposed an exact lens equation and integral formulation for a  spherically symmetric BH. An analytical logarithmic expansion technique to calculate the strong gravitational lensing for a spherically symmetric BH was proposed in \cite{Bozza2002} and corresponding observables lens equations for the BH were obtained. Ref. \cite{Perlick} also discussed the various gravitational lensing observables for static and spherically symmetric BH and it has been used to study the different type of BHs such as RNBH \cite{Eiroa},  BHs with string theory \cite{Bhadra}, Horndeski BH \cite{Horn}, rotating regular BHs \cite{Jusufi}, RNdS BH \cite{Zhao}, Kerr and Kerr-Newman BHs \cite{Hsieh2021a, Hsieh2021b}, GUP-modified Schwarzschild BH \cite{Turakhonov}, charged BH with global monopole \cite{lan2025}, Kerr-de Sitter BH \cite{Omwoyo}, and a holonomy-corrected Schwarzschild BH \cite{Junior}. Ref. \cite{Bozza2004b} explore the time delay between various relativistic images produced by  strong field gravitational lensing around a static, spherically symmetric BH.


One basic symmetry in spacetime is Lorentz symmetry. It serves as the cornerstone of both  the quantum field theory and the standard model of contemporary particle physics, and is also essential to general relativity (GR). However, the detection of high-energy cosmic ray signals \cite{Takeda} and the development of unified canonical theories indicate that spontaneous Lorentz symmetry breaking might take place at higher energy scales.  Generally, effective field theory can be used to characterize Lorentz violation effects, which are only empirically observable at sufficiently low energy scales \cite{Casana2018, Kostelecky1995}. The bumblebee gravity model provides a straightforward and useful classical field theory framework for investigating Lorentz symmetry breaking \cite{Kostelecky2004,Bluhm}. By introducing the bumblebee vector field $\beta_\mu$ with a non-zero vacuum expectation value (VEV), this model induces spontaneous Lorentz symmetry breaking, thereby altering the symmetric nature of the background spacetime. Thus, the bumblebee gravity framework can reveal novel physical phenomena which cause it foundation for the emergence of modern physics. The bumblebee gravitational model was first introduced by Kostelecký and Samuel \cite{Kostelecky1989} to investigate the effects of spontaneous Lorentz violation. A Schwarzschild-like bumblebee BH was later presented by Casana et al. \cite{Casana2018}. Researchers then identified various spherical solutions including wormhole geometries \cite{Ovgun2019}, cosmological constants \cite{Maluf,Priyo2022,Oni2022,Media2023,Oni2023,Media2025}, global monopoles \cite{Gullu, Belchior} and Einstein-Gauss-Bonnet terms \cite{Ding2022}. The framework was later extended by Ding et al. \cite{Ding2020} obtaining a Kerr-like rotating solution.

The study of perturbations in BH geometry plays a significant role in gravitational physics. Regge and Wheeler \cite{Regge} pioneered BH perturbation theory by examining metric perturbations of the Schwarzschild BH and demonstrated that the radial equation for axial perturbations resembles the Schrödinger equation. Since then, many researchers have extended from their findings and derived similar equations for various BHs with different spins. Vishveshwara \cite{Vish} and Chandrasekhar \cite{Chandra} explained the understanding of BH perturbations through their work on quasinormal modes (QNMs). The QNMs of a BH describe its natural oscillations, with complex-valued frequencies related to the solutions of the perturbation solution that hold the special boundary conditions for purely ingoing waves near the event horizon and outgoing waves at spatial infinity. The real part and the imaginary part of QNMs correspond to the oscillation frequencies of the perturbation and the exponential decay rate due to energy loss through gravitational waves, respectively. Field perturbations around a BH   generally occurs in three distinct stages \cite{Frolov}. The initial stage corresponds to an initial burst of radiation, governed by the nature of the initial perturbation. This is followed by a phase of damped oscillations defined as QNMs in which frequencies and damping times are fully governed  by the geometry of the BH spacetime and remain unaffected by the initial perturbation.
 The final stage occurs at a very late times, where the waves exhibit a power-law decay due to backscattering effects from the surrounding gravitational field. 
  Since the detection of gravitational waves by LIGO and Virgo \cite{Abbott}, several methods have been proposed for finding QNMs of various kinds of BHs such as Wentzel-Kramers-Brillouin (WKB) approximative method \cite{Schutz1985,Iyer1987a,Iyer1987b,Wahlang,Roshila,Jayasri}, Frobenius method \cite{Konoplya2011}, Poschl-Teller fitting method \cite{Ferrari,PriyoEPJC,MediaGRG}, Continued fraction method (Leaver's method) \cite{Leaver}, Mashhoon method \cite{Mashhoon} and Asymptotic Iteration method (AIM) \cite{Cho2009,Cho2010,Pong2019,Pong2020,Gogoi}. Additionally, GF plays a fundamental role in understanding wave propagation and energy emission in perturbed BH spacetimes. It quantifies the transmission probability of waves,  indicating how much of the wave is transmitted to infinity or absorbed by the event horizon  \cite{Konoplya2019,Cardoso,Dey}. Different methods have used to analyse the GF such as  rigorous bound method \cite{Visser,Sakalli,Boonserm2008,Ngam2013a,Ngam2013b,Boonserm2014a, Boonserm2014b, Boonserm2018,Badawi}, WKB approximation method for high gravitational potential \cite{Parikh2000,Konoplya2020}, matching method \cite{Fernando,Kim},  and analytical method for different spin field \cite{Cardos2003, Pano,Rincon}.

We organize this paper as follows:  In section 2, we explore the  gravitational lensing effects under strong field limit and derive the lensing observables for different values of $L$ and $Q$. We discuss the lensing effects of SMBHs and  constraint  the parameter $L$ from the EHT shadow obseravations in sections 3 and 4 respectively. In section 5, we discuss the Dirac equation in Newman-Penrose formalism and the corresponding effective potential is derived. Section 6 is devoted to the calculation of QNMs by using sixth-order WKB method and AIM method. Section 7 explores the potential for detecting QNMs of RNdS-like BH. In section 8, we calculate the GF for different values of $L$ and $Q$. The absorption cross sections and the sparsity of Hawking radiation are also discussed in section 9 and section 10 respectively. The findings of the paper are presented in section 11.
\section{Strong lensing of charged BH in bumblebee gravity}
\begin{figure*}[!htbp]
\centering
\subfloat[  ]
{\includegraphics[width=175pt,height=155pt]{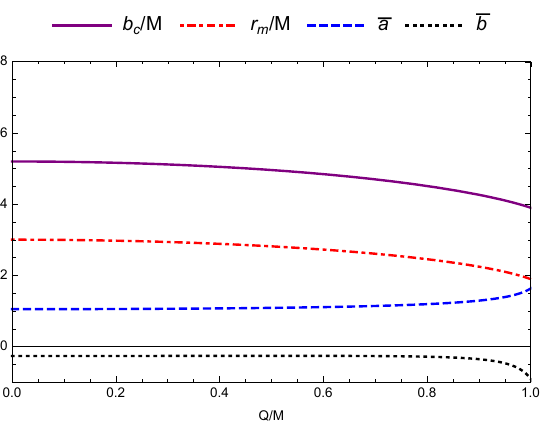}
\label {g1}
}
\hfill
\subfloat[]
{\includegraphics[width=175pt,height=155pt]{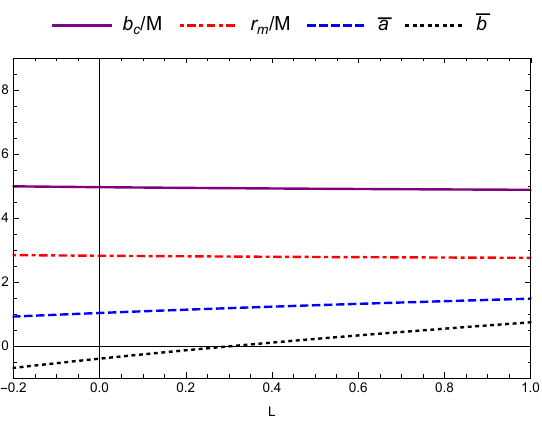}
\label {g2}
}
\caption{Plot of $b_c/M$, $r_m/M$, $\bar{a}$ and $\bar{b}$ in RN-like BH as a function of  (a) $Q/M$ with fixed $L=0.2$ and (b) $L$ with fixed $Q=0.5M$.}
\label{G1}
\end{figure*}

\begin{figure*}[!htbp]
\centering
\subfloat[]
{\includegraphics[width=175pt,height=155pt]{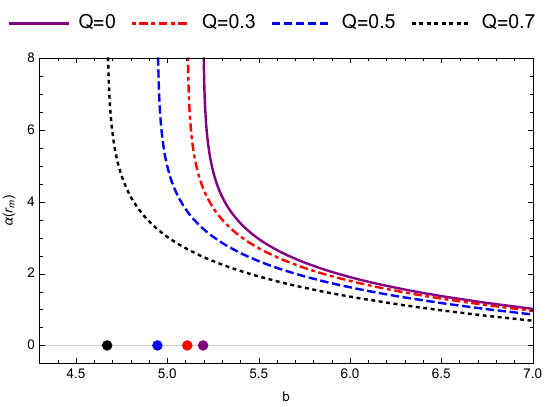}
\label {g3}
}
\hfill
\subfloat[]
{\includegraphics[width=175pt,height=155pt]{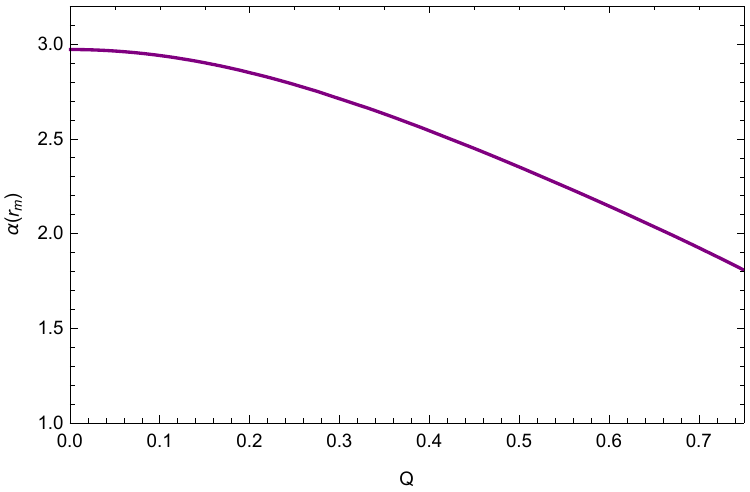}
\label {g4}
}
\caption{(a) The deflection angle $\alpha(r_m)$ in the strong field limit as a function of impact parameter $b$ for varying $Q$ with fixed $L=0.2$. The divergent points of the deflection angle are shown with the coloured points on the horizontal axis. (b) $\alpha(r_m)$ with $Q$ for fixed $b=5.5$.}
\label{G2}
\end{figure*}

\begin{figure*}[!htbp]
\centering
\subfloat[]
{\includegraphics[width=175pt,height=155pt]{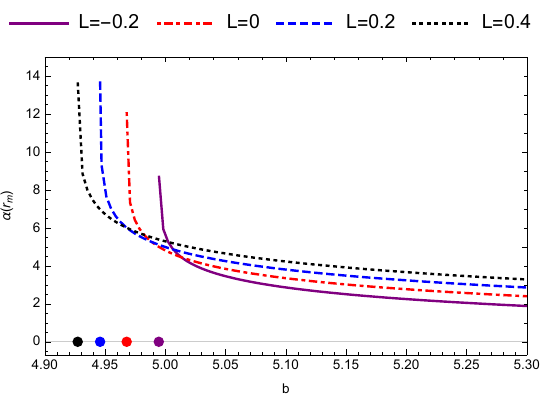}
\label {g5}
}
\hfill
\subfloat[]
{\includegraphics[width=175pt,height=155pt]{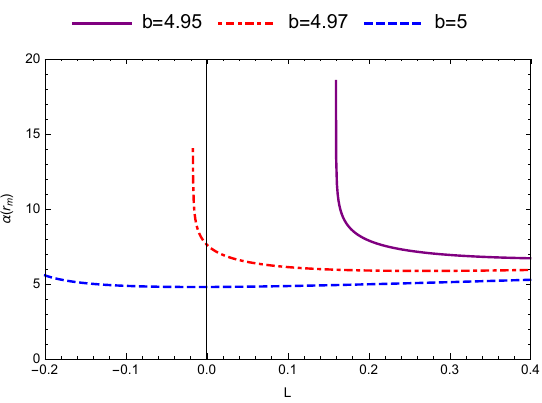}
\label {g6}
}
\caption{(a) The deflection angle $\alpha(r_m)$ in the strong field limit as a function of impact parameter $b$ for different values of $L$ with fixed $Q=0.5M$. The coloured points on the horizontal axis are the divergent points of the deflection angle. (b) $\alpha(r_m)$ with $L$ for different values of $b$.}
\label{G3}
\end{figure*}

In this section, we first consider the spacetime geometry described by the line element of the RNdS-like BH in bumblebee gravity, given by \cite{Liu2025}

\begin{align}\label{le}
ds^2=A(r) dt^2-B(r) dr^2-C(r) \left(  \sin^2{\theta} ~d\phi^2 +d\theta^2 \right),
\end{align}
where
\begin{align}\label{e3}
&A(r)=1-\frac{2M}{r}-\frac{\Lambda r^2 (1+L)}{3}+\frac{2Q^2(1+L)}{r^2(2+L)},\cr 
&B(r)=\dfrac{(1+L)}{A(r)}, \,\,
C(r)=r^2.
\end{align}
We know that Eq. \eqref{le} is asymptotically flat space for $\Lambda=0$. Therefore, the strong lensing of \eqref{le} will be investigated only for RN-like BH. Since the spacetime is a static and spherical symmetric, there exist time translational $t^\mu \partial_\mu=\partial_t$ and axial Killing vectors $\phi^\mu \partial_\mu=\partial_\phi$, respectively.
The  event horizon located at $r=r_H$ is obtained from $A(r)=0$. Outside the event horizon there exists a spherical region which is known as photon sphere on which the  photon orbits the BH along circular paths. Since the circular  photon orbits are not stable,  a  small perturbation may cause the photon to fall into the BH or escape to infinity.  The  photon sphere radius $r_m$ represents the minimum distance at which a photon can stably orbit the BH  and  is determined as  the largest root of the equation

\begin{align}
A(r)C^{'}(r)-A'(r) C(r)=0,
\end{align}
where prime denotes the first derivative with respect to  $r$. Therefore, the photon sphere $r_m$ is derived as
\begin{align}\label{rm}
r_m=\frac{3M+\sqrt{9M^2-\frac{16Q^2(1+L)}{2+L}}}{2}.
\end{align}
It is observed from the above equation that the photon sphere $r_m$ depends not only on BH mass $M$, charge $Q$ but also on the $L$. The real photon sphere $r_m$ can be achieved only when $|\frac{M}{Q}|>\frac{4}{3}\sqrt{\frac{1+L}{2+L}}$. We can calculate the path of a light ray  from the equation 
\begin{align}\label{tra}
g_{\mu \nu} k^\mu k^\nu=0.
\end{align}
Here $k^\mu=\dot{x}^\mu$ is the wave number of the photon and the dot represents derivative  with respect to an affine parameter of the null geodesic respectively. The photon's  angular momentum  $\mathcal{L}$ and  energy $E$   are constants along the null geodesic and are defined as
\begin{align}
&E\equiv -g_{\mu \nu} t^\mu k^\nu=A(r) \dot{t}, \hspace{0.5cm} \mathcal{L} \equiv g_{\mu \nu} \phi^\mu k^\nu=C(r) \dot{\phi}.
\end{align}
It is noted that $E$ and $\mathcal{L}$ are non zero. For a light ray, the impact parameter $b$ is defined by
\begin{align}
b\equiv \frac{\mathcal{L}}{E}=\frac{C(r) \dot{\phi}}{A(r) \dot{t}}.
\end{align}
For simplicity, both $\mathcal{L}$ and $b$ are taken as non-negative for a single light ray and  motion is restricted to the equatorial plane. By using Eq. \eqref{tra}, we derive the trajectory of photon in the RN-like BH as 
\begin{align}\label{tr}
A(r)\dot{t}^2-B(r)\dot{r}^2-C(r)\dot{\phi}^2=0
\end{align}
or
\begin{align}\label{ra}
\dot{r}^2=V_{eff}(r),
\end{align}
where $V_{eff}(r)$ is the effective potential defined by
\begin{align}\label{ef}
V_{eff}(r) \equiv \dfrac{1}{1+L} \left(E^2-\frac{\mathcal{L}^2 A}{ r^2}\right).
\end{align} 
The photon motion is allowed only in the region where $V_{eff}$ remains positive. A photon may reach at the spatial infinity since the effective potential $V_{eff}(r)$ tends to $ \frac{E^2}{1+L}>0$ as $r \rightarrow \infty$. We assume that a photon which comes from the infinity  with  impact parameter $b$, toward the BH is symmetrically deflected back to infinity after reaching its closest approach at  $r=r_0$ near the BH. At the closest distance $r=r_0$, the radial velocity $\dot{r}$ tends to zero. From Eqs. \eqref{ra} and \eqref{ef}, the impact parameter $b$ is related to the closest distance $r_0$ and can be expressed as \cite{Bozza2002}
\begin{align}\label{b1}
b(r_0)=\sqrt{\frac{C_0}{A_0}}.
\end{align}
Here the subscript $0$ signifies that the quantity is  evaluated at $r=r_0$. The critical impact parameter $b_c$ is calculated using  Eqs. \eqref{e3} and \eqref{b1} as
\begin{align}
b_c(r_m)&= \lim_{r_0 \to r_m}\sqrt{\frac{C_0}{A_0}}\cr
&= \frac{r_m^2}{\sqrt{M r_m-\frac{2Q^2(1+L)}{2+L}}}.
\end{align}
The deflection angle can be  represented as an expansion  of $r_0-r_m$ with respect to  $r=r_0$ in the spacetime, defined by
\begin{align}\label{b2}
b(r_0)&=b_c(r_m)+\frac{1}{4} \sqrt{\frac{C_m}{A_m}} (r_0-r_m)^2 \Big(\frac{C_m^{''}}{C_m}-\frac{A_m^{''}}{A_m} \Big)\cr
&+O(r_0-r_m)^3.
\end{align}
Differentiating Eq. \eqref{ef} with respect to the radial coordinate $r$, we obtain
\begin{align}
V'_{eff}(r)=2\mathcal{L}^2 \left(\frac{1}{r^3}-\dfrac{3M}{r^4}+\dfrac{4Q^2(1+L)}{(2+L )r^5}\right).
\end{align}
Using Eqs. \eqref{rm} and \eqref{ef}, we get
\begin{align}
 \lim_{r_0 \to r_m} V_{eff}(r_0)= \lim_{r_0 \to r_m} V_{eff}^{'}(r_0)=0.
\end{align}
Thus, when the photon's impact parameter $b$ approaches its critical value $ b_c$,  it almost stops moving radially just  outside the light sphere at radius $r=r_m$. In the strong deflection limit the photon's orbit winds around the light sphere as $r_0 \to r_m$ or equivalently $b \rightarrow b_c$. The trajectory Eq. \eqref{tr} becomes
\begin{eqnarray}
\Big(\frac{dr}{d\phi} \Big)^2&=-\frac{2Q^2}{2+L}\Big(\frac{r}{r_m}\Big)^4+\frac{M r^4}{(1+L)r_m^3}-\frac{r(r-2M)}{1+L}-\frac{2Q^2}{2+L}. \nonumber\\
\end{eqnarray}
The deflection angle of the light ray $\alpha(r_0)$ is the angle between the asymptotic outgoing and incoming trajectories and is expresses as
\begin{align}
\alpha(r_0)=I(r_0)-\pi,
\end{align}
with $I(r_0)$ given by
\begin{align}\label{de1}
I(r_0)\equiv \int_{r_0}^{\infty} \frac{2~dr}{\sqrt{-\frac{2Q^2}{2+L}\Big(\frac{r}{r_m}\Big)^4+\frac{M r^4}{(1+L)r_m^3}-\frac{r(r-2M)}{1+L}-\frac{2Q^2}{2+L}}}.
\end{align}
We can introduces a new variable $x$ as \cite{Tsukamoto2017b}
\begin{align}
x \equiv 1-\frac{r_0}{r}.
\end{align}
Then the deflection angle from Eq. \eqref{de1} becomes
\begin{align}
I(r_0)=\int_{0}^{1} f(x,r_0) dx,
\end{align}
where
\begin{align}
f(x,r_0)=\frac{2r_0}{\sqrt{c_1(r_0)x+c_2(r_0)x^2+c_3(r_0)x^3+c_4(r_0)x^4}}.
\end{align}
Here, the values of $c_n(r_0)$, $n=1,2,3,4$, are given as 
\begin{align}
&c_1(r_0)=\dfrac{2}{1+L}\left(r_0^2-3M r_0+\frac{4Q^2 (1+L)}{2+L}\right),\cr
&c_2(r_0)=\frac{3M r_0}{1+L}-\frac{8Q^2}{2+L}, \cr
&c_3(r_0)=-2\left(\frac{M r_0}{1+L}-\frac{4Q^2}{2+L}\right),\,\,\,
c_4=-\frac{2Q^2}{2+L}.
\end{align}
In the strong deflection limit, $r_0 \to r_m$, the coefficient $c_1(r_m)$ vanishes and  $c_2(r_m)$ takes the value
\begin{align}
c_2(r_m)=\frac{3M r_m}{1+L}-\frac{8Q^2}{2+L}.
\end{align}
This implies $f(x,r_0)$ diverges with order  $x^{-1}$.
The term $I(r_0)$ can be expressed in terms of a regular part $I_R(r_0)$ and a divergent part $I_D(r_0)$  as:
\begin{align}
I(r_0)=I_R(r_0)+I_D(r_0).
\end{align}
The divergent part is given by
\begin{align}\label{I1}
I_D(r_0)=\int_{0}^{1} f_D(x,r_0) dx,
\end{align}
where
\begin{align}
f_D(x,r_0)=\frac{2r_0}{\sqrt{c_1(r_0)x+c_2(r_0)x^2}}.
\end{align}
Then Eq. \eqref{I1} becomes
\begin{align}
&I_D(r_0)\cr
&=\frac{4r_0}{\sqrt{\frac{3M r_0}{1+L}-\frac{8Q^2}{2+L}}} \log \left[\frac{\sqrt{\frac{3M r_0}{1+L}-\frac{8Q^2}{2+L}}+\sqrt{\frac{2r_0^2}{1+L}-\frac{3M r_0}{1+L}}}{\sqrt{2(\frac{r_0^2}{1+L}-\frac{3M r_0}{1+L}+\frac{4Q^2}{2+L})}}\right]. \nonumber\\
\end{align}
Using Eq. \eqref{b2} and taking the strong deflection limit $r_0 \to r_m$ or $b \to b_c$, $I_D(r_0)$ can be written  as
\begin{align}
I_D(r_m)=&-\bar{a} \log \Big(\frac{b}{b_c}-1 \Big)+\bar{a} \log  X +O((b-b_c) \cr
&\times \log(b-b_c)),
\end{align}
where
\begin{align}
\bar{a}&=\frac{r_m\sqrt{(2+L)(1+L)}}{\sqrt{3M r_m(2+L)-8Q^2(1+L)}},\cr
X&=\frac{2(3M r_m(2+L)-8Q^2(1+L))}{Y},\cr
Y&=Z \times \left\lbrace 3M r_m-\frac{4Q^2(1+L)}{2+L}\right\rbrace^2,\cr
Z&=M r_m(2+L)-2Q^2(1+L).
\end{align}
The regular part is given by
\begin{align}
I_R(r_0)\equiv \int_{0}^{1} f_R(x,r_0) dx,
\end{align}
where
\begin{align}
f_R(x,r_0)\equiv f(x,r_0)-f_D(x,r_0).
\end{align}
For the deflection angle in the strong deflection limit $r_0 \to r_m$, we have
\begin{align}
 \lim_{r_0 \rightarrow r_m}f_R(x,r_0)&= \frac{2r_m}{x\sqrt{c_2(r_m)+c_3(r_m) x+c_3(r_m) x^2}}\cr
 &-\frac{2r_m}{x\sqrt{c_2(r_m)}} .
\end{align}
In the strong deflection limit $r_0 \rightarrow r_m$ or $b_0 \rightarrow b_c$, an analytical expression can be derived as
\begin{align}
I_R(r_m)&= \bar{a} \,\, \log \Big[ \frac{X^2 Y^2}{M^2 r_m^2(2+L)^2Z}\cr
& \times \Big(2\sqrt{M r_m(2+L)-2Q^2(1+L)}\cr
&-\sqrt{\frac{X Y}{2}} \Big)^2 \Big] +O((b-b_c)\log(b-b_c)).
\end{align}
The deflection angle $\alpha(r_m)$ in the strong deflection limit $b_0 \rightarrow b_c$ is defined by
\begin{align}\label{q1}
\alpha(r_m)=-\bar{a}\,\, \log \Big(\frac{b}{b_c}-1 \Big)+\bar{b}+O((b-b_c)\log(b-b_c)),
\end{align}
where
\begin{align}
\bar{b}&=\bar{a}\,\, \log X+I_R(r_m)-\pi.
\end{align}
It is noted that $\bar{a}$ is the positive function which depends on   $L$, charge $Q$ and BH mass $M$. The function $\bar{b}$ may be negative or positive depending  upon the choice of  $L$. When $L$ tends to zero, we obtain $b_c$, $r_m$, $\bar{a}$ and $\bar{b}$ of RNBH \cite{Tsukamoto2017b}. The values of $b_c$, $r_m$, $\bar{a}$ and $\bar{b}$ are consistent with the values of Schwarzschild BH \cite{Tsukamoto2017b} when $Q=L=0$. Figs. \ref{g1} and \ref{g2} show the graphs of $b_c/M$, $r_m/M$, $\bar{a}$ and $\bar{b}$ as a function of $Q/M$ and $L$ respectively. We see that increasing $Q/M$ and $L$ deviate the values of $\bar{a}$ and $\bar{b}$ from the Schwarzschild BH as shown in Figs. \ref{g1} and \ref{g2}. The radius of photon sphere also decreases with the increase of $Q$ and $L$ but becomes imaginary for $|Q|>\frac{3M}{4}\sqrt{\frac{2+L}{1+L}}$ and $L>\frac{18M^2-16Q^2}{16Q^2-9M^2}$. If the BH has no charge $(Q=0)$, we recover the Schwarzschild-like BH  and corresponding values of $b_c$, $r_m$, $\bar{a}$ and $\bar{b}$ are given by
\begin{align*}
&r_m=3M,\,\, b_c=3\sqrt{3}M,\,\, \bar{a}=\sqrt{1+L},\cr
& \bar{b}=\sqrt{1+L}\,\, \rm log \left[1512-864\sqrt{3}\right]-\pi.
\end{align*}
The graphs of deflection angle $\alpha(r_m)$ in the strong deflection limit for the RN-like BH are shown in Figs. \ref{G2} and \ref{G3}. It is noted that the deflection angle $\alpha(r_m)$ is divergent for $b=b_c$ and suddenly falls with $b$ in Fig. \ref{G2}. For fixed values of $b$, the deflection angle $\alpha(r_m)$ decreases with increasing $Q$. We also see that the deflection angle of Schwarzschild BH in bumblebee gravity is larger than RN-like BH. In Fig. \ref{G3}, one can see that the deflection angle suddenly decreases with increasing $L$ when $b$ is small,  whereas for larger values of  $b$ and $L$ it increases monotonically. It is also noted that when $b$ is small, the deflection angle of RNBH is bigger than the RN-like BH.
Since the strong field limit holds if $b$ tends to $b_c$, any valid results can't be derived if $b$ is bigger than $b_c$. Hence, there are some values of $x_0=x_y$ or $b=b_y$ such that the deflection angle approaches zero.

\subsection{Lens observation}
In this section, we examine how the parameters $L$ and $Q$ affect  the observables in the strong field limit. To derive a simple and reliable expression for the deflection angle, containing both a logarithmic term and a constant term, we shall apply the strong field limit approximation. Using Eq. \eqref{q1} into the lens equation, we can derive a direct connection between the position and the magnification of the images and the deflection angle which are derived in accordance with the strong field limit. In the strong field limit, the lens equation is given by \cite{Bozza2001} 
\begin{align}\label{l1}
\beta=\theta-\frac{D_{LS}}{D_{OS}} \Delta \alpha_n,
\end{align}
where $\theta$ and $\beta$ denote the angular separation of the lens from the image and the source from the lens. $D_{OL}$  and   $D_{LS}$ denote the distances between the observer and the lens, and between the lens and the source, respectively. The total observer–source distance is $D_{OS}=D_{OL}+D_{LS}$.
We assume that the light rays complete $n$ revolution around the lens before reaching the observer in the strong field limit and $\Delta \alpha_n=\alpha(\theta)-2n \pi$ gives the deviation of the deflection angle. 
For deriving the position for the $n^{th}$ relativistic image $\Delta \alpha_n$, we firstly calculate $\alpha_D(\theta_n^0)=2n\pi$, where $\theta_n^0$ represents the corresponding image position,  which gives
\begin{align}\label{l2}
\theta_n^0=\frac{b_c}{D_{OL}}(1+e_n),
\end{align}
where $e_n=e^{\bar{b}-2n\pi/\bar{a}}$.
Expanding the deflection angle about $\theta_n^0$ to the first order by using  the first order Taylor series expansion, the deflection angle about $\theta_n^0$ is obtained as
\begin{align}\label{l3}
\alpha_D(\theta)=\alpha_D(\theta_n^0)+\frac{\partial \alpha_D(\theta)}{\partial \theta}\Big|_{\theta_n^0} (\theta-\theta_n^0)+O(\theta-\theta_n^0),
\end{align}
where
\begin{align}\label{l4}
\Delta \alpha_n=-\frac{\bar{a} D_{OL}}{b_c e_n}\Delta \theta_n.
\end{align}
 The position of $n^{th}$ image can be derived from the lens equation by ignoring the higher-order terms as \cite{Bozza2002, Molla}
\begin{align}\label{l5}
\theta_n \simeq \theta_n^0+\frac{D_{OS}}{D_{LS}} \frac{b_c e_n}{D_{OL} \bar{a}} (\beta-\theta_n^0).
\end{align}
If $\beta=\theta_n^0$, then the correction to the $n^{th}$ image position will be zero i.e. the source position coincides with image position. It is to be noted that Eq. \eqref{l5} represents source image on the same side of source $\theta>0$. The image on the other side of the source can be obtained if we replace $\beta$ by $-\beta$.

We know that the gravitational lensing effect magnifies the brightness of the source's image and corresponding magnification of the $n^{th}$ relativistic image is given by \cite{Bozza2002}
\begin{align}\label{l6}
\mu_n=\Big(\frac{\beta}{\theta} \frac{d\beta}{d\theta} \Big|_{\theta_n^0} \Big)^{-1}.
\end{align}
We also get
\begin{align}\label{l7}
\frac{d\beta}{d\theta}=1+\frac{\bar{a} D_{OL}}{b_c e_n}\frac{D_{LS}}{D_{OS}}.
\end{align}
The first term  can be neglected since it is sufficiently small compared to the second. Therefore, the magnification of $n^{th}$ image takes the form
\begin{align}\label{l8}
\mu_n=e_n \frac{b_c^2  D_{OS}(1+e_n)}{\bar{a} \beta D_{OL}^2 D_{LS}}.
\end{align}
It is noted from Eq. \eqref{l8} that $\mu_n$ is inversely proportional to $D_{OL}^2$ so that a bright image can be achieved only when $\beta \to 0$. Hence the brightness of the relativistic images becomes faint and decreases with the increase of $n$. We know that the first relativistic images becomes brightest and its brightness decreases quickly. Eqs. \eqref{l2} and \eqref{l8} give the relationship between the position and magnification to strong field limit coefficients which also carry the information about the nature of BH. The relationship between the relativistic image observables and the strong field limit coefficients can be derived by using the deflection angle formulas and the lens equation. The three essential observables are
(i) the asymptotic position $ (\theta_\infty)$ approached by a set of images, obtained from  Eq. \eqref{l5} as $n \to \infty \,$
(ii) the angular separation $(s)$ between the first image and the asymptotic position $\theta_\infty$ and 
(iii) the ratio of the flux $(r_{mag})$ of the first image to the combined flux of all other images  which are given by
\begin{align}\label{l9}
& \theta_{\infty}=\frac{b_c}{D_{OL}}, \cr
& s=\theta_{\infty} e^{\frac{\bar{b}-2\pi}{\bar{a}}}, \cr
& r=e^{\frac{2\pi}{\bar{a}}},\,\,\,
r_{\text{ mag}}= \frac{5\pi}{\bar{a}\,\, \text{ ln} 10}.
\end{align}
The above expressions show that the observables are completely determined by the strong-field limit coefficients, the critical impact parameter and observer-lens distance. Now  we extend our analysis to Einstein ring and time delay, which are also important observational feature of strong gravitational lensing.

\subsection{Einstein ring}
The Einstein ring is formed when a massive foreground object  such as a BH or galaxy, perfectly aligns with a source, lens and the observer (i.e. $\beta=0$) \cite{Tsukamoto2017b}. This occurs due to gravitational lensing in which the object’s immense gravity distorts spacetime, bending light from the background source to curve around it. When $\beta=0$, Eq. \eqref{l5} can be written as
\begin{align}\label{er1}
\theta^{E}_n=\left(1-\frac{D_{OL}}{D_{LS}}\frac{b_c e_n}{D_{OL} \bar{a}}\right)\theta^0_n .
\end{align}
 Assuming perfect alignment of the source, observer and  lens where the lens  located at the midpoint between the source and the observer ($D_{OS}=2D_{OL}$), the angular radius of Einstein ring can be obtained using Eq. \eqref{l2} in Eq. \eqref{er1} as 
\begin{align}\label{ringsnew}
\theta^{E}_n=\left(1-\frac{2  e_n b_c }{ \bar{a} D_{OL} }\right) \times \frac{(1+e_n)b_c}{D_{OL}}.
\end{align}
Since $D_{OL}\gg b_c$, the angular radius of the $n^{th}$ relativistic Einstein
 ring can be obtained as
\begin{align}\label{er2}
\theta^{E}_n=\frac{(1+e_n) b_c}{D_{OL}}.
\end{align}
When $n=1$ the above expression represents the angular radius of the outermost Einstein ring. From Eq. \eqref{er2}, we also observe that the angular radius of the Einstein ring decreases when the distance between observer and lens increases and increases for large critical impact parameter $b_c$. 

\subsection{Time delay}
\begin{figure*}[!htbp]
\centering
\subfloat[]
{\includegraphics[width=175pt,height=155pt]{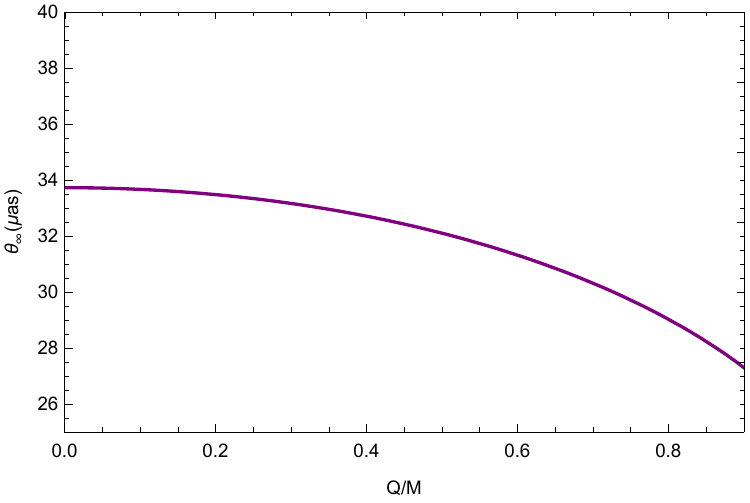}
\label {g7}
}
\hfill
\subfloat[]
{\includegraphics[width=175pt,height=155pt]{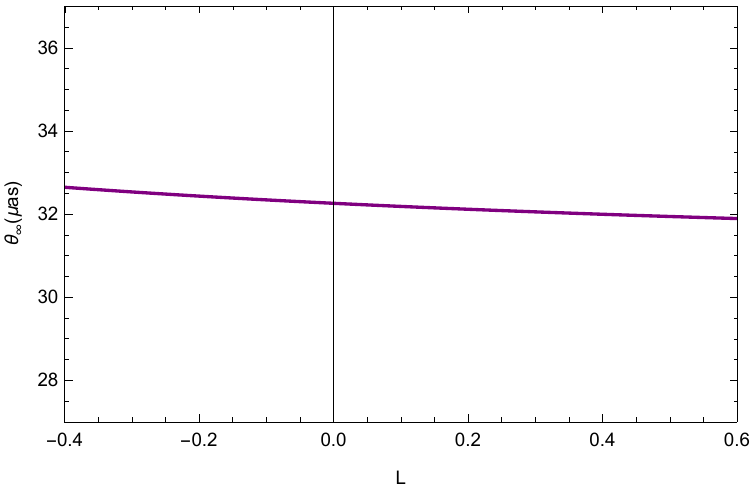}
\label {g8}
}
\caption{Plot of lensing observables $\theta_\infty$ as a function of (a) $Q/M$ with $L=0.2$ and (b)  $L$ with $Q=0.5M$.}
\label{G4}
\end{figure*}

\begin{figure*}[!htbp]
\centering
\subfloat[]
{\includegraphics[width=175pt,height=155pt]{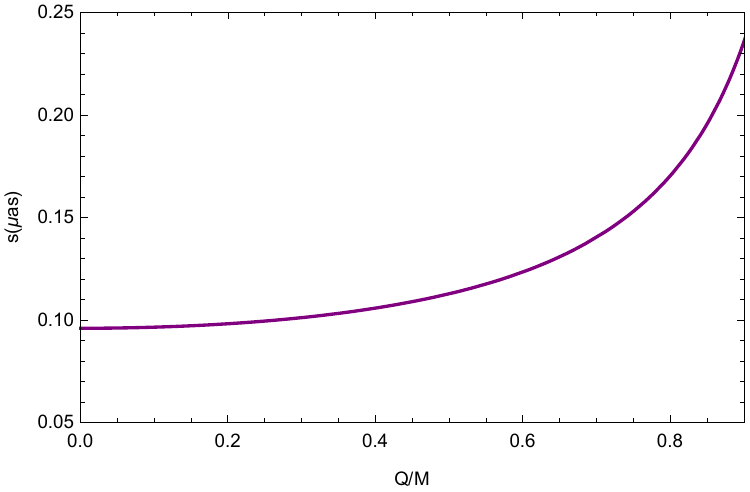}
\label {g9}
}
\hfill
\subfloat[]
{\includegraphics[width=175pt,height=155pt]{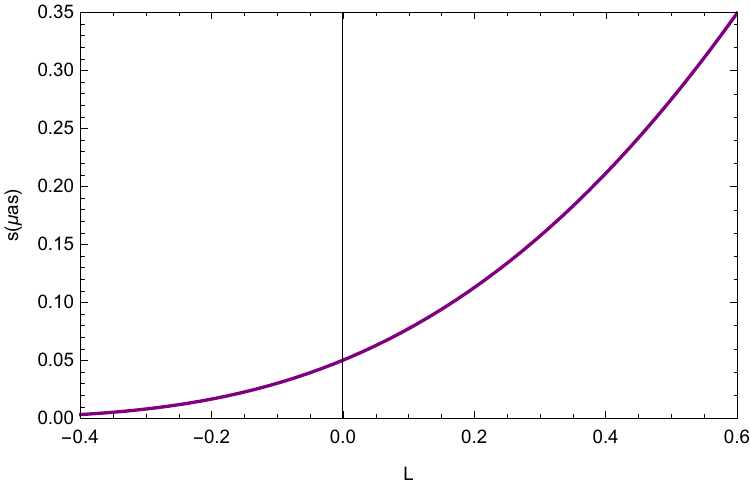}
\label {g10}
}
\caption{Plot of lensing observables $s$ (a)varying $Q/M$ for $L=0.2$ and (b) varying $L$ for $Q=0.5M$.}
\label{G5}
\end{figure*}

\begin{figure*}[!htbp]
\centering
\subfloat[]
{\includegraphics[width=175pt,height=155pt]{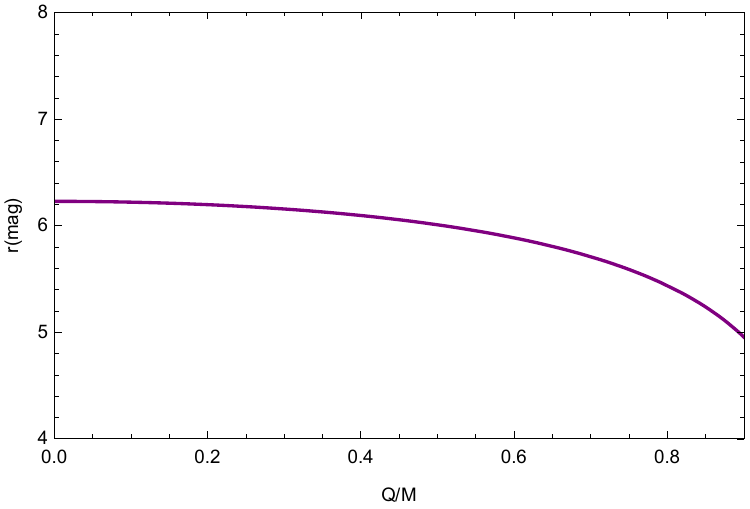}
\label {g11}
}
\hfill
\subfloat[]
{\includegraphics[width=175pt,height=155pt]{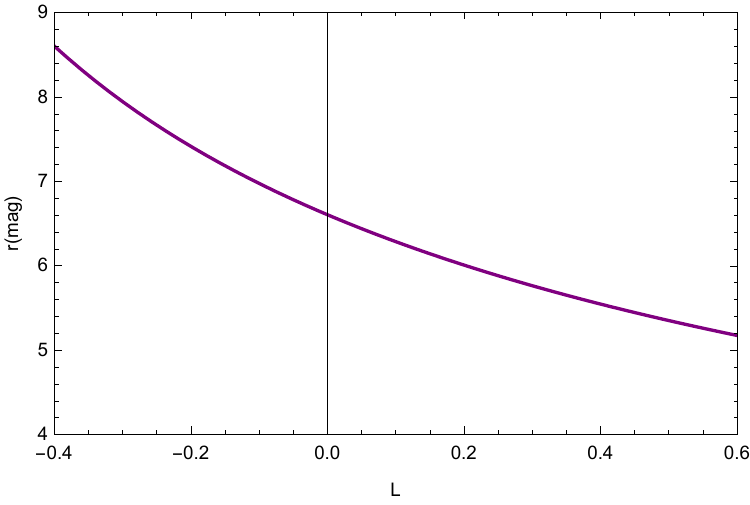}
\label {g12}
}
\caption{Plot of the relativistic image $r_{mag}$ (a) varying $Q/M$ for $L=0.2$ and (b) varying $L$ for $Q=0.5M$.}
\label{G6}
\end{figure*}

\begin{figure*}[!htbp]
\centering
\subfloat[]
{\includegraphics[width=175pt,height=155pt]{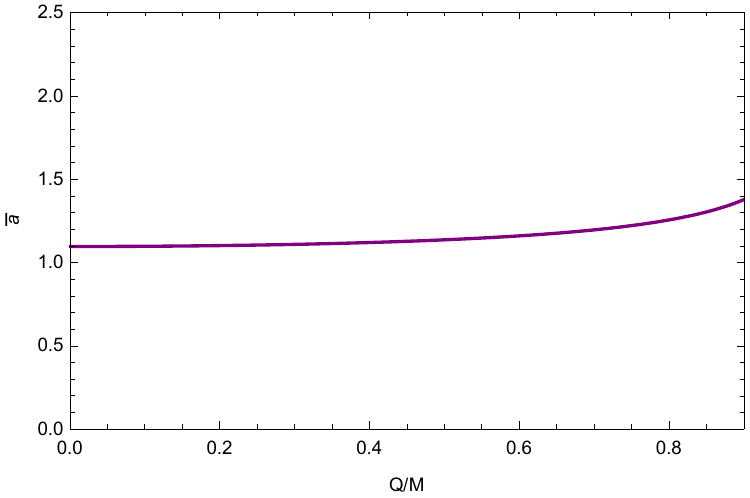}
\label {g13}
}
\hfill
\subfloat[]
{\includegraphics[width=175pt,height=155pt]{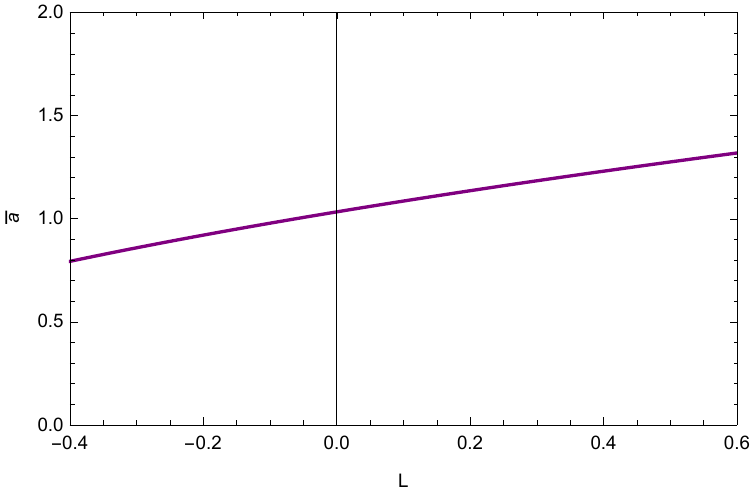}
\label {g14}
}
\caption{Plot of the lensing coefficient $\bar{a}$ (a) varying $Q/M$ for $L=0.2$ and (b) varying $L$ for $Q=0.5M$.}
\label{G7}
\end{figure*}

\begin{figure*}[!htbp]
\centering
\subfloat[]
{\includegraphics[width=175pt,height=155pt]{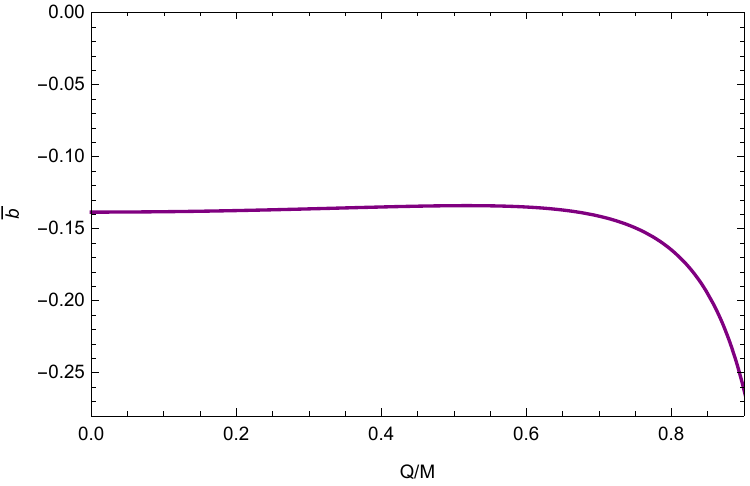}
\label {g15}
}
\hfill
\subfloat[]
{\includegraphics[width=175pt,height=155pt]{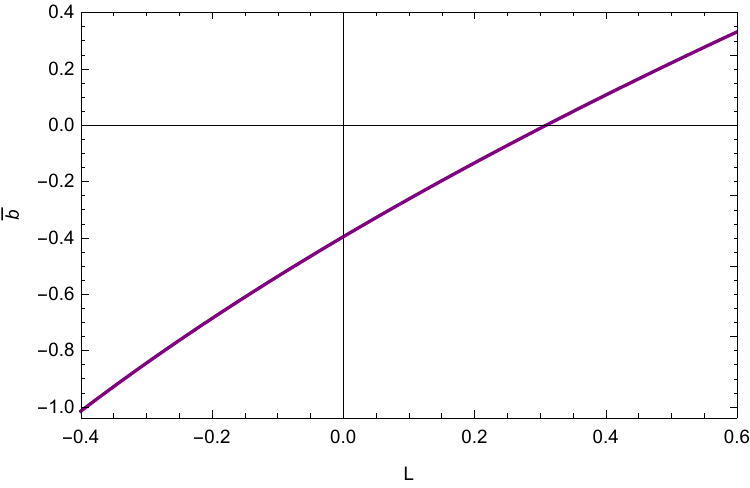}
\label {g16}
}
\caption{Plot of the lensing coefficient $\bar{b}$ (a) varying $Q/M$ for $L=0.2$ and (b) varying $L$ for $Q=0.5M$.}
\label{G8}
\end{figure*}

\begin{figure*}[!htbp]
\centering
\subfloat[$L=0.2$]
{\includegraphics[width=175pt,height=155pt]{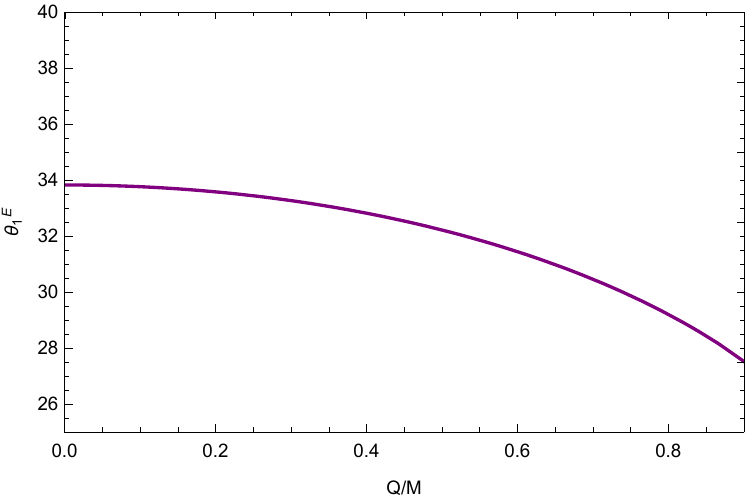}
\label {eringQ}
}
\hfill
\subfloat[$Q=0.5M$]
{\includegraphics[width=175pt,height=155pt]{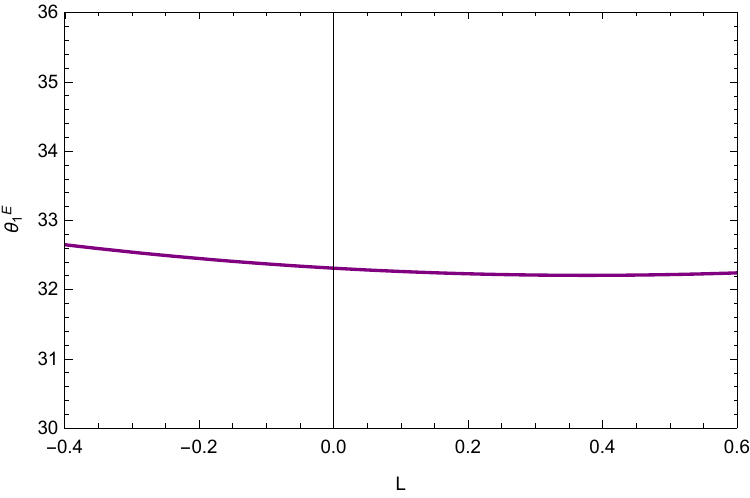}
\label {eringL}
}
\caption{Plot of the Einstein ring $\theta^E_1$ (a)varying $Q/M$ for $L=0.2$ and (b) varying $L$ for $Q=0.5M$.}
\label{ERING}
\end{figure*}

\begin{figure*}[!htbp]
\centering
\subfloat[]
{\includegraphics[width=175pt,height=155pt]{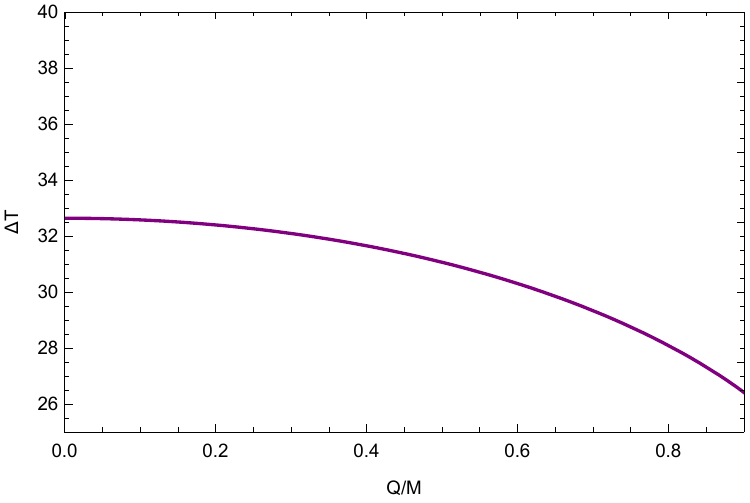}
\label {g17}
}
\hfill
\subfloat[]
{\includegraphics[width=175pt,height=155pt]{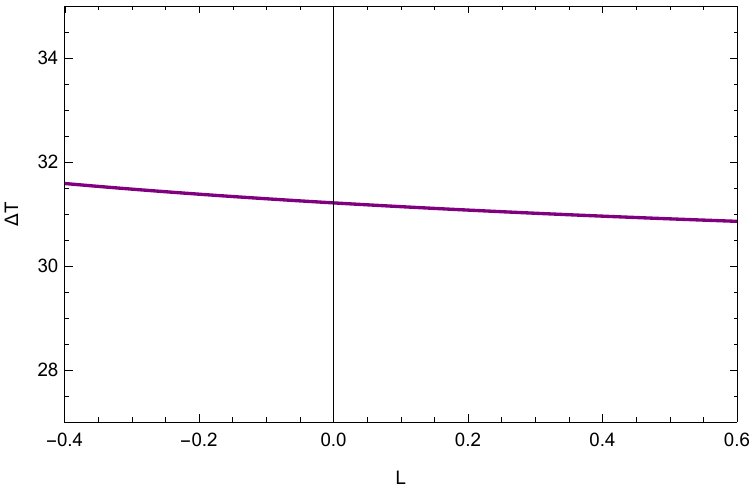}
\label {g18}
}
\caption{Plot of the lensing observables $\Delta T$ (a)varying $Q/M$ for $L=0.2$ and (b) varying $L$ for $Q=0.5M$.}
\label{G9}
\end{figure*}

In the strong-field lensing, the time delay arises when photons take different trajectories around the BH, leading to variations in time travel and resulting in an arrival-time difference between two relativistic images.
By comparing the  time signals of two relativistic images, the time delay between them can be determined \cite{Molla}. The duration for a photon to complete one circular orbit around the BH is given by \cite{Molla}
\begin{align}\label{time delay 1}
T(\tilde{b})=\bar{a} \log \left(\dfrac{b}{b_c}-1\right) +\bar{b}+O\left(b-b_c\right).
\end{align}
For the relativistic images positioned on the same side of the lens, the time delay between the first and the second  is given by
\begin{align}
\Delta T^{s}_{2,1}=2 \pi b_c=2\pi D_{OL} \theta_\infty.
\end{align}
For the RN-like BH, the lensing observables $\theta_\infty$, $s$ and $r_{mag}$ with $Q/M$ and $L$ are drawn in Figs. \ref{G4}, \ref{G5} and \ref{G6} respectively. It is noted that both the observables $\theta_\infty$ and $r_{mag}$ weakly decrease for higher values of $Q/M$ and $L$. However, the observable $s$ increases with increasing of $Q/M$ and $L$. The lensing coefficients $\bar{a}$ and $\bar{b}$ are illustrated in Figs. \ref{G7} and \ref{G8}. It is evident that $\bar{a}$ and $\bar{b}$ increase for larger values of $Q/M$ and $L$. The angular radius of the outermost Einstein ring $\theta^E_n$ with $Q/M$ and $L$ are dipicted in Fig. \ref{ERING}. It shows that $\theta^E_n$ slowly decreases when the values of $Q/M$ and $L$ increase. The time delay $\Delta T^s_{2,1}$ between the first and the second relativistic images for RN-like BH are plotted as a function of $Q/M$ and $L$  in Figs. \ref{g17} and \ref{g18} respectively. It shows that the $\Delta T^s_{2,1}$ weakly decreases with increasing $Q/M$ and $L$. The validation of the above statements are also shown numerically in Tables \ref{T1} and \ref{T2} respectively.

\begin{table*}[htp]
\caption{Computed lensing observables and strong field limit coefficients
 for varying $Q/M$ with fixed parameters $L=0.2$.}
\label{T1}
\begin{tabular}{p{1cm} p{1.7cm} p{1.7cm} p{1.7cm} p{1.7cm} p{1.7cm} p{1.7cm} p{1.5cm}}
\toprule
$\frac{Q}{M}$ & $\theta_{\infty}(\mu a s) $ &  $s(\mu a s)$  & $r_m(mag)$ &  \,\,\,\,$\bar{a}$& \,\,\,\,\,\,$\bar{b}$& $\Delta T_{2,1}^s$ & \,\, $\theta^E_1$ \\ \midrule
0   & 33.7400  & 0.0960 & 6.2275 &  1.0955 & -0.1386 & 32.6484 & 33.836 \\
0.1 & 33.6785  & 0.0965 & 6.2199 &  1.0968 & -0.1383 & 32.5889 & 33.775 \\
0.3 & 33.1769  & 0.1012 & 6.1560 &  1.1082 & -0.1364 & 32.1035 & 33.2781 \\ 
0.5 & 32.1147  & 0.1128 & 6.0079 &  1.1355 & -0.1341 & 31.0757 & 32.2275 \\
0.7 & 30.3284  & 0.1403 & 5.7083 &  1.1951 & -0.1413 & 29.3472 & 30.4687 \\
0.9 & 27.3163  & 0.2355 & 4.9552 &  1.3767 & -0.2613 & 26.4325 & 27.5517 \\ \bottomrule
\end{tabular}
\end{table*}

\begin{table*}[htp]
\caption{Computed lensing observables and strong field limit coefficients
 for varying $L$ with fixed $Q=0.5M$.}
\label{T2}
\begin{tabular}{p{1cm} p{1.7cm} p{1.7cm} p{1.7cm} p{1.7cm} p{1.7cm} p{1.7cm} p{1.5cm}}
\toprule
$L$ & $\theta_{\infty}(\mu a s) $ &  \,$s(\mu a s)$  & $r_m(mag)$ & \,\,\,\,$\bar{a}$& \,\,\,\,\,\,$\bar{b}$&  $\Delta T_{2,1}^s$ & \,\, $\theta^E_1$ \\ \midrule
-0.4 & 32.6440 & 0.0033 & 8.6041  &  0.7929 & -1.0155 & 31.5878  & 32.6473  \\
-0.2 & 32.4310 & 0.0166 &  7.4145 &  0.9201 & -0.6863 &  31.3817 & 32.4476  \\
0   & 32.258 & 0.0501 &  6.6044   &  1.0329 & -0.3963 & 31.2143  & 32.3081  \\ 
0.2 & 32.1147 & 0.1128 &  6.0079  &  1.1355 & -0.1341 & 31.0757  & 32.2275  \\
0.4 & 31.9941 & 0.2112 &  5.5455  &  1.2302 & 0.1069 & 30.9589   & 32.2052  \\
0.6 & 31.8911 & 0.3493 & 5.1738   &  1.3186 & 0.3311 &  30.8593  & 32.2404  \\ \bottomrule
\end{tabular}
\end{table*}

\section{Lensing effects of the supermassive black holes}
\begin{figure*}[!htbp]
\centering
\subfloat[]
{\includegraphics[width=175pt,height=155pt]{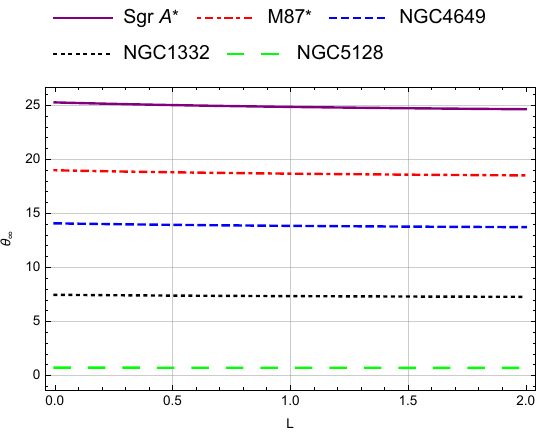}
\label{thetainfL}
}
\hfill
\subfloat[]
{
\includegraphics[width=175pt,height=155pt]{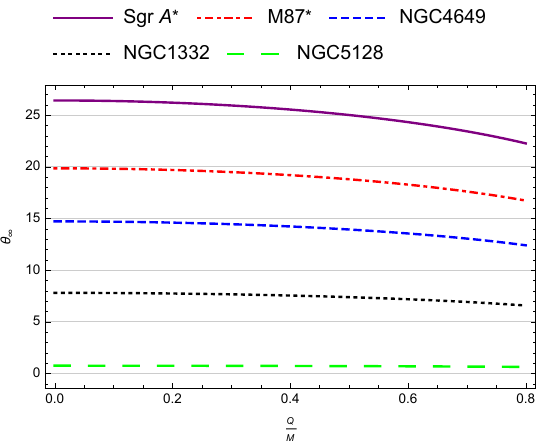}
\label{thetainfQ}
}
\caption{ Variation of angular image position $\theta_\infty (\mu as)$ for Sgr A*, M87*, NGC 5128, NGC 1332 and  NGC 4649 for varying (a) $L$ with fixed $Q/M=0.5$ and (b) $Q/M$ with fixed $L=0.5$  in strong field limit.}
\label{thetainf}
\end{figure*}

\begin{figure*}[!htbp]
\centering
\subfloat[]
{\includegraphics[width=175pt,height=155pt]{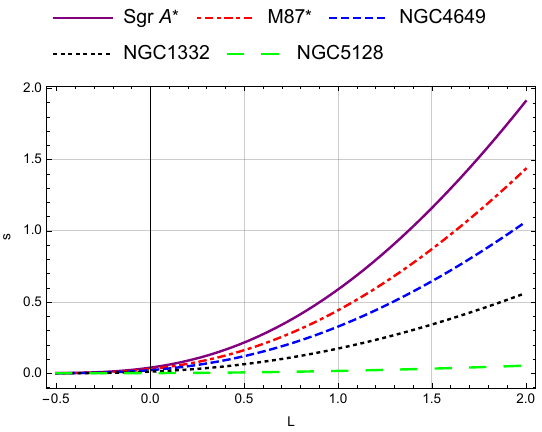}
\label {sinf}
}
\hfill
\subfloat[]{
\includegraphics[width=175pt,height=155pt]{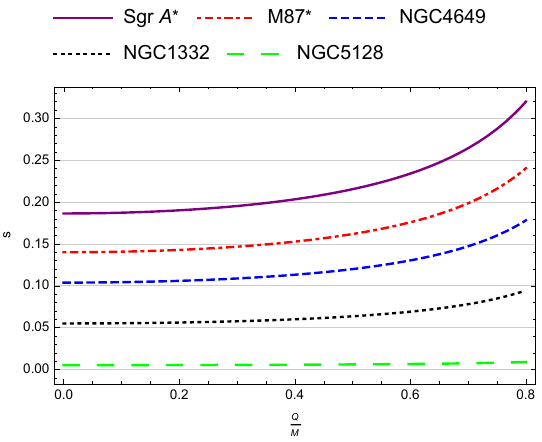}
\label{sinfQ}
}
\caption{The angular image separation $s(\mu as)$ for Sgr A*, M87*, NGC 5128, NGC 1332 and  NGC 4649 in strong field limit for varying (a) $L$ when $Q/M=0.5$ is fixed and (b) $Q/M$ when $L=0.5$.}
\label{fig:}
\end{figure*}

In this section, the strong gravitational lensing quantities $\theta_\infty$, $s$, $r_{mag}$, $\Delta T^{s}_{2,1}$ and Einstein ring for  RN-like BH are estimated by using data from SMBHs such as Sgr A*, M87*, NGC 5128, NGC 1332, NGC 4649 etc.  These results are compared with the corresponding observables calculated for the RNBH. The mass and distances from Earth for several SMBHs are as follows: Sgr A*  has a mass of 
$4.3 \times 10^6 M_\odot$   and lies at a distance of 8.35 kpc \cite{Akiyama2022a}; M87*  has a mass of $6.5 \times 10^9 M_\odot$ and lies 16.8 Mpc away \cite{Akiyama2022b}; NGC 4649  has a mass of $4.72 \times 10^9M_\odot$ and is positioned at 16.46 Mpc  from Earth \cite{Kormendy}; and NGC 1332  possesses $2.54 \times 10^9 M_\odot$, situated at 16.72 Mpc from Earth \cite{Kormendy}; NGC 5128  has a mass of $5.69 \times 10^7 M_\odot$ and is located at 3.62 Mpc away \cite{Harris}.
The observational values of mass and distance  are converted into geometrized units for consistency with the theoretical framework.  The dependence of  $\theta_\infty$ on the parameters $L$ and $Q/M$ for various SMBHs are shown in Figs. \ref{thetainfL} and \ref{thetainfQ} respectively. As $L$ and $Q/M$ increase, we observe a gradual decrease in the $\theta_\infty$ of all the SMBHs, although the rate and the magnitude of change vary slightly depending upon the BH's mass and distance. This indicates that increasing  $L$ and $Q/M$ cause the light rays to be deflected less strongly, thereby reducing the apparent angular size of the photon sphere observed by a distant observer.  The effect of $L$ is more noticeable in the less massive BH, Sgr A*.
Figs. \ref{sinf} and \ref{sinfQ} illustrate the  dependence  of the strong lensing observable $s$ on $L$ and $Q$ respectively. Increasing $L$ and $Q$ cause   to increase the gravitational observable  $s$ for all BHs. Thus  the angular spread of relativistic images gets higher with increasing $L$ and $Q/M$. Table \ref{tabobservation1} presents the estimated lensing observables $\theta_\infty$, $s$ and $r_{mag}$  for the RN-like BH model, evaluated for the SMBHs  M87*, Sgr A*, NGC 5128, NGC 1332 and NGC 4649 . The numerical values presented in Table \ref{tabobservation1} are consistent with the findings  in Figs. \ref{sinf} and \ref{sinfQ}, further validating the monotonic behaviour of $\theta_\infty$
  with  both  $L$ and $Q/M$. Moreover, the value of $r_{mag }$ remains same  for all the BHs considered, implying that it is determined by the intrinsic properties of the BH rather than their mass or distance. Further increasing $L$ and $Q/M$, $r_{mag}$ decreases, thereby making the inner relativistic images become comparatively brighter and more observable. The estimated time delay $\Delta T_{2,1}^s$ between the first and the second relativistic images for SMBHs located at the centers of various galaxies are presented in Table \ref{observation2}. The case with $L=0$ corresponds to the standard RNBH.  Moreover, for BH of the same mass and distance, our analysis finds the time delay of standard RNBH is higher than the RN-like BH in bumblebee gravity. For all  types of BHs, the time delay decreases with increasing $L$ and $Q$. This indicates that Lorentz violation modifies the effective spacetime geometry and a higher Lorentz violation parameter $L$ allows the photon to complete their orbits around the BH in a shorter duration. Further, the time delay is significantly large for high-mass BHs such as M87* and NGC 4649 when  compared to less massive BHs like Sgr A* and NGC 5128. Since the time delay for Sgr A* is around 10 minutes, its observational detection becomes particularly challenging due to the short temporal separation between relativistic images.

\begin{figure}[!htbp]
\centering
{\includegraphics[width=175pt,height=175pt]{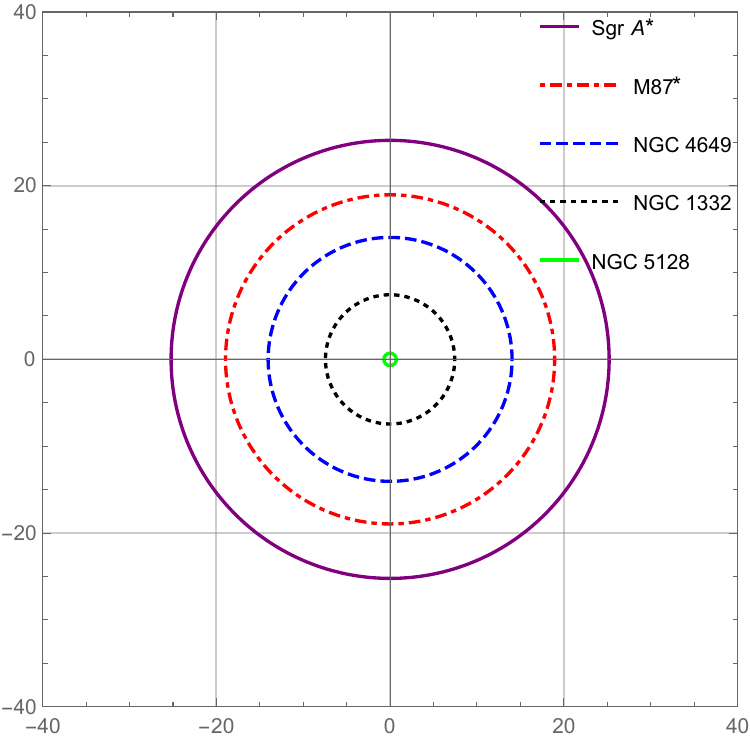}
}
\caption{Plot of Einstein rings of  Sgr A$^*$, M87$^*$, NGC 4649, NGC 1332 and NGC 5128 for $L=0.5$ and $Q=0.5M$.}
\label{Ering}
\end{figure}

The outermost relativistic Einstein rings of SMBHs Sgr A$^*$, M87$^*$, NGC 5128, NGC 1332 and  NGC 4649 are plotted in Fig. \ref{Ering}. It can be seen that the angular radius of Einstein rings of Sgr A$^*$ is the greatest and NGC 5128 is the smallest.

\section{Constraints from the EHT observation}

\begin{figure}[!htbp]
\centering
\subfloat[]
{\includegraphics[width=175pt,height=155pt]{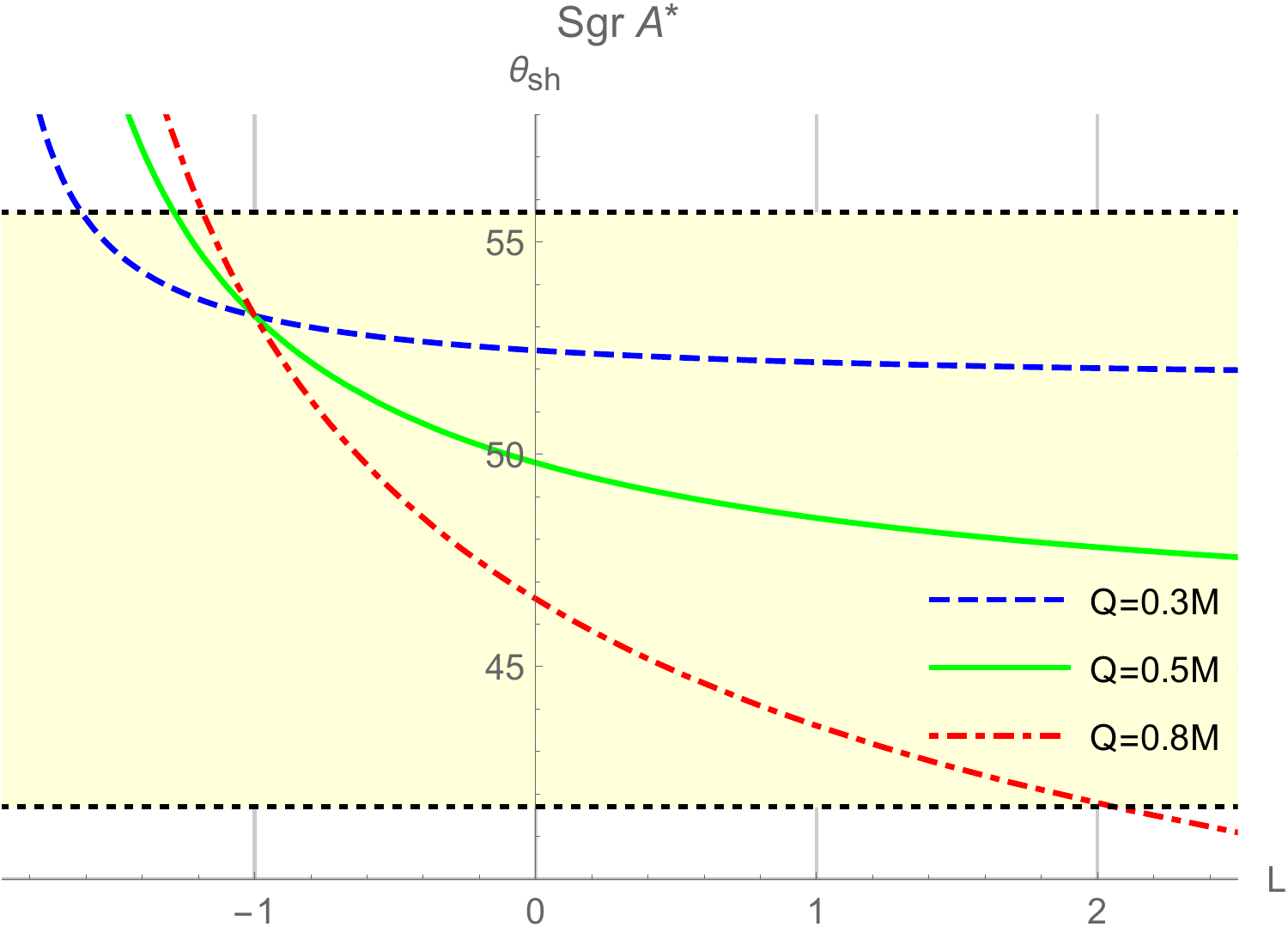}
\label {SgrA}
}
\hfill
\subfloat[]
{\includegraphics[width=175pt,height=155pt]{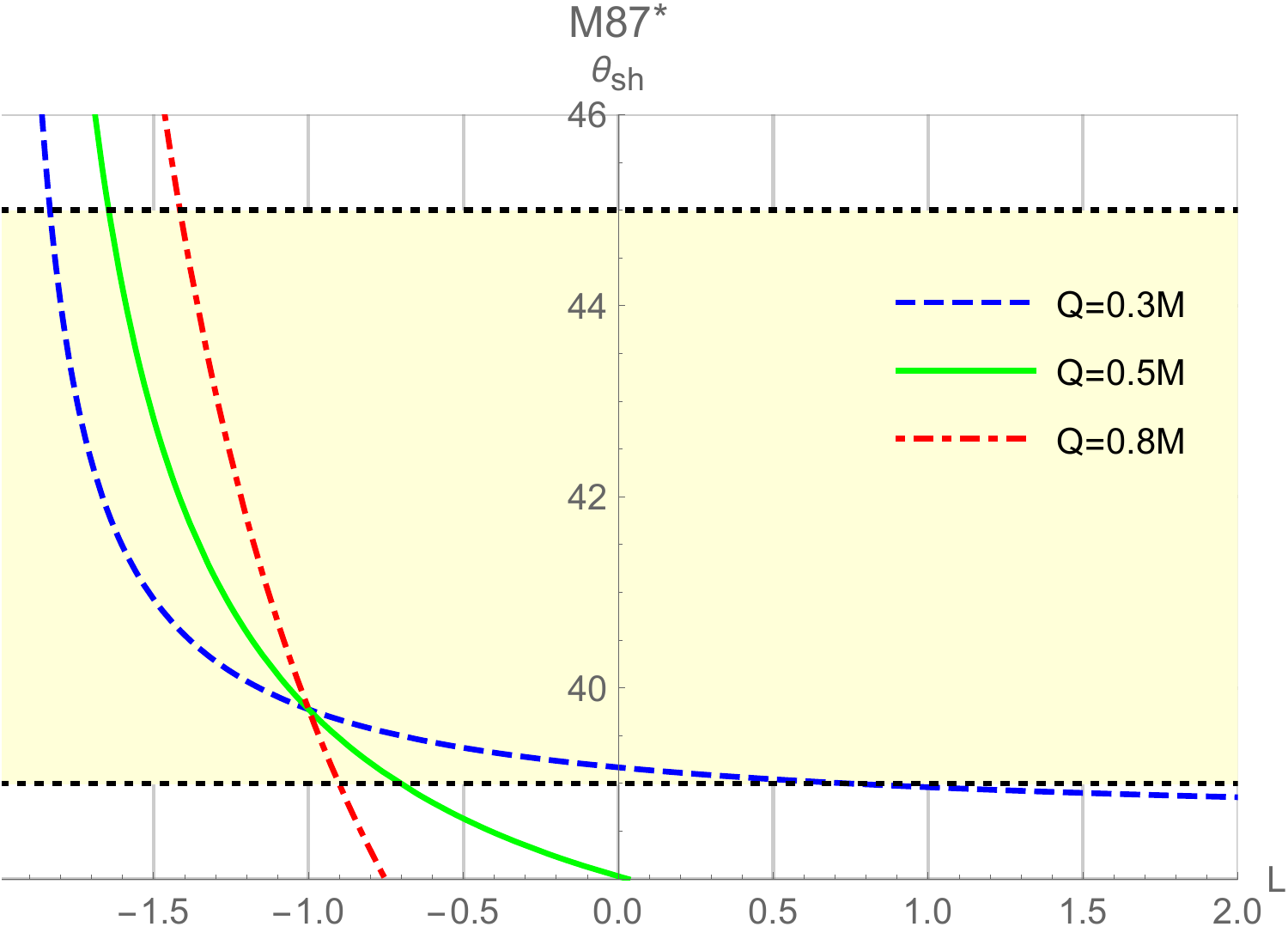}
\label {M87}
}
\caption{Constraints on the parameter $L$ from the 1$\sigma$ observed shadow angular diameters of (a) Sgr A$^*$ and (b) M87$^*$. The shaded horizontal bands indicate the observational 1$\sigma$ ranges.}
\label{constraint}
\end{figure}

The investigation of BH shadows has become a powerful tool for exploring the strong-field regime of gravity.  By analyzing the shape and size of the shadow, one can test the predictions of general relativity and explore deviations arising from alternative gravity theories. Following the breakthrough achievement of Event Horizon Telescope (EHT) in capturing the BH shadows of M87* and Sgr A*, extensive efforts have been made to constrain BH parameters and explore alternative theories of gravity. Using the EHT shadow data, constraints on the  parameter $L$ have previously been explored in  \cite{wang2022,wang2022b,islam2024}.  Our analysis extends this approach by systematically comparing the theoretical predictions of shadow structure  with the EHT shadow observations for Sgr A* and M87*. Refs. \cite{Akiyama2022a,Akiyama2022b} report that the angular diameter of the Sgr A*  and M87* are  $\theta_{\text{Sgr A*}}=48.7\pm 7\mu as$ and $\theta_{\text{M87*}}=42\pm 3 \mu as$ respectively.  In Fig. \ref{constraint}, we illustrate the constraints placed on the parameter $L$ for various values of $Q$ using the EHT data of 1$\sigma$ observational limits on the angular shadow diameter of Sgr A* and M87*. It is clear that the allowed range of $L$ depends  on $Q$. The numerical constraints derived from the figure are summarized in Table \ref{tabconstraint}.

\begin{sidewaystable}[p]
\centering
\vspace{8cm}
\captionof{table}{Estimates for the lensing observables for RN-like BH in the context of SMBHs  Sgr A*, M87*, NGC 4649 and NGC 1332 for different values of the $Q$ and $L$.}\label{tabobservation1}
\begin{tabular}{llllllllllllll}
\hline
 &   & \multicolumn{2}{c}{Sgr A*} & \multicolumn{2}{c}{M87*} & \multicolumn{2}{c}{NGC 4649} & \multicolumn{2}{c}{NGC 1332} & \multicolumn{2}{c}{NGC 5128} & \\ 
$Q$ &  $L$  &   $\theta_{\infty} (\mu \text{as})$        &     $s (\mu \text{as})$      &     $\theta_{\infty} (\mu \text{as})$        &    $s (\mu \text{as})$            &      $\theta_{\infty} (\mu \text{as})$       &  $s (\mu \text{as})$              &       $\theta_{\infty} (\mu \text{as})$      &     $s (\mu \text{as})$   &      $\theta_{\infty} (\mu \text{as})$       &  $s (\mu \text{as})$           &  $r_{mag}$ \\ \hline
0.5M &   -0.3     & 25.4659   &  0.00631649  & 19.1329  & 0.00474567    & 14.1804    &  0.00351727   &  7.44914   &  0.0115808  & 0.715742			&	0.00017753	& 	6.60439\\

 &   0     & 25.2517   &  0.0392576  & 18.972   & 0.0294948    & 14.0612    &  0.0218602   &  7.51234   &  0.00186334  & 0.709721	&	0.00110337	&   6.60439\\
	 &   0.3   &  25.0904  &  0.123054  & 18.8508   &  0.0924525   &  13.9713    &  0.0685215   &   7.40155  &  0.0363004  &0.705187	&	0.00345855	&  5.76315\\
	 &   0.6   &  24.9645  &  0.273454  &  18.7562  &   0.20545  &  13.9012    &  0.15227   &  7.36441   &  0.0806678  & 0.701649	& 0.00768568		&   5.17376\\
	 &   0.9   &  24.8635  &  0.497009  &  18.6803  & 0.37341    &   13.845   &  0.276754   &  7.33461   & 0.146615   & 0.698809	&	0.0139689	&   4.73196\\ \hline

	 0.8M &  -0.3     &  23.7844  & 0.00946705   & 17.8696   & 0.00711274    &   13.2441    &    0.00527163 &  7.0163   &  0.00279274  & 0.668482	&	0.00026608	&  6.07378 \\

 &   0     &  23.1071  & 0.0594693   & 17.3607   & 0.0446802    &   12.8669    &    0.0331149 &  6.81648   &  0.0175432  & 0.649444	&	0.00167144	&  6.07378 \\
	 &   0.3   &  22.5649  &  0.184972  &  16.9534  &   0.138972  &  12.565   &    0.103  & 6.65655    & 0.054566   & 0.634207 	&	0.0051988	& 5.17461 \\
	 &   0.6   &  22.1174  &  0.403122  & 16.6171   & 0.302872    &   12.3158   &  0.224475   &  6.52453  &  0.118919  & 0.621628	&	0.0113301	&  4.53683 \\
	 &   0.9   &  21.739  &  0.712612  & 16.3328   &  0.535396   &  12.1051   &  0.39681    & 6.4129    & 0.210217   & 0.610993	&	0.0200286	& 4.05311  \\
\end{tabular}
\end{sidewaystable}

\begin{sidewaystable}[p]
\centering
\vspace{7cm}
\captionof{table}{Estimates and comparison of time delay for SMBHs at the center of  different galaxies between RNBH and RN-like BH in bumblebee gravity. Time delays are expressed in minutes. }\label{observation2}
\begin{tabular}{llllllll| lll l}
Galaxy & $M(M_{0})$  & $D_{OL}(Mpc)$  & $M/D_{OL}$ &  $\Delta T_{2,1}^s $ & $\Delta T_{2,1}^s $ & $\Delta T_{2,1}^s $ & $\Delta T_{2,1}^s $ & $\Delta T_{2,1}^s $ &	$\Delta T_{2,1}^s $\\ \hline
 &  &  &  & \multicolumn{3}{c|}{Q=0.5M} & \multicolumn{3}{c}{Q=0.8M} \\ \cline{5-10} 
&	&	&	& L=-0.3	& L=0	& L=0.3	&L=0.6	&	L=-0.3 & L=0& L=0.3 & L=0.6	\\ \cline{5-10}
 Sgr A* & $4.3 \times 10^6$ & 0.00835 & $2.46429\times10^{-11}$ &	 11.1044	& 11.011  & 10.9407 & 10.8858 &	 	10.3712  & 10.0758  & 9.83943 & 9.64428 \\
 M87*& $6.5 \times 10^9$ & 16.8  & $1.85146\times10^{-11}$ &	16785.7 	 & 16644.5 & 16538.2 & 16455.2 &	15677.4 	 & 15230.9	& 14873.6 & 14578.6 \\
 NGC 4649 & $4.72 \times 10^9$ & 16.46 & $1.37221\times10^{-11}$  &	 12189	& 12086.5 & 12009.3 & 11949 &	 11384.2	 & 11060 & 10800.5	& 10586.3\\
NGC 1332 & $2.54 \times 10^9$ &  16.72& $7.26953\times10^{-12}$ &	 6559.35 	 & 6504.17 & 6462.62 & 6430.19  &	 6126.24	&  5951.77 & 5812.13	& 5696.85\\
NGC 5128 & $5.5 \times 10^7$ & 3.8 & $6.92609\times10^{-13}$ &	142.033 	  & 140.838 & 139.939 & 139.236  &	132.655 	 & 128.877 & 125.853 & 123.357 \\
NGC 4697 &$2.02 \times 10^8$  & 12.54 & $7.70838\times10^{-13}$ &	 521.649	  & 517.261 & 513.957 & 511.378 &	 487.205 	 & 473.33 & 462.224	&453.057 \\
 NGC 4374&  $9.25 \times 10^8$& 18.51 & $2.39136\times10^{-12}$ &	 2388.74	 & 2368.65 & 2353.51 & 2341.7 &	 2231.01	 & 2167.48 & 2116.62 & 2074.64\\
NGC 3608 & $4.65 \times 10^8$ & 22.75 & $9.78094\times10^{-13}$ &	 1200.83	 & 1190.72 & 1183.12 & 1177.18 &	 1121.54	 & 1089.6 &	1064.03& 1042.93 
\end{tabular}
\end{sidewaystable}

\begin{table}[htbp]
\centering
\caption{Comparison of constraints on the  parameter $L$ derived from the EHT angular shadow size for Sgr A* and M87* at various values of charge-to-mass ratio $Q/M$.}
\begin{tabular}{ccccc}
\hline
\textbf{$Q/M$} & \textbf{Source} & \textbf{$\theta_{\text{sh}}^{\text{obs}}$ [$\mu$as]} & \textbf{ $L_{\text{min}}$} & \textbf{ $L_{\text{max}}$} \\
\hline
\multirow{2}{*}{0.3} 
& Sgr A* & $41.7 - 55.7$ & $-1.61674$ & $---$ \\
& M87*  & $39 - 45$ & $-1.83459$ & $0.735942$ \\
\hline
\multirow{2}{*}{0.5} 
& Sgr A* & $41.7 - 55.7$ & $-1.36681$ & $---$ \\
& M87*  & $39 - 45$ & $-1.64494$ & $-0.703961$ \\
\hline
\multirow{2}{*}{0.8} 
& Sgr A* & $41.7 - 55.7$ & $-1.18453$ & $2.05516$ \\
& M87*  & $39 - 45$ & $-1.41505$ & $-0.902033$ \\
\hline
\end{tabular}
\label{tabconstraint}
\end{table}


\section{Chandrasekhar Dirac equation in NP formalism}
This section focuses on the study of massless Dirac perturbations in the background of RNdS-like BH. Using the NP formalism, we will discuss the wave equation for the RNdS-like BH. The four coupled Dirac equations written in NP formalism are defined by
\begin{align}\label{D1}
&(D+\varepsilon-\rho+iq l^\mu A_\mu)F_1+(\overline{\delta}+\pi-\alpha+iq\overline{m}^\mu A_\mu)F_2\cr
&=i\mu^*G_1,\nonumber\\
&(\delta+\beta-\tau+iqm^\mu A_\mu)F_1+(\Delta+\mu-\gamma+iqn^\mu A_\mu)F_2\cr
&=i\mu^*G_2,\nonumber\\
&(D+\overline{\varepsilon}-\overline{\rho}+iq l^\mu A_\mu)G_2-(\delta+\overline{\pi}-\overline{\alpha}+iq m^\mu A_\mu)G_1\cr
&=i\mu^*F_2,\nonumber\\
&(\Delta+\overline{\mu}-\overline{\gamma}+iq n^\mu A_\mu)G_1-(\overline{\delta}+\overline{\beta}-\overline{\tau}+iq m^\mu A_\mu)G_2\cr
&=i\mu^*F_1.
\end{align}
Here $D,\Delta,\delta,$ and $\overline{\delta}$ represent the directional derivatives and $\alpha,\beta,\rho,\varepsilon,\mu,\pi,\gamma $ and $\tau$ represent spin coefficients  and $\mu^*$ denote the mass parameter of the Dirac spinor in NP formalism \cite{Newman}. In the NP formalism, the Dirac spinor is represented by four components $F_1, F_2, G_1$ and $G_2$. 
The covariant and the contravariant form of the null tetrad vectors for spherically symmetric  BH are given by
\begin{align}&l_i= \left( 1,\dfrac{-\sqrt{1+L}}{A},0,0 \right),\nonumber\hspace{0.2cm}
n_i=\left(\dfrac{A}{2},\dfrac{\sqrt{1+L}}{2},0,0 \right),\\
&m_i=\left(0,0,\dfrac{-r}{\sqrt{2}},\dfrac{-ir\sin\theta}{\sqrt{2}}\right),\,\,
\overline{m}_i=\left(0,0,\dfrac{-r}{\sqrt{2}},\dfrac{ir\sin\theta}{\sqrt{2}}\right),
\end{align}
and
\begin{align}
&l^i= \left( A^{-1},\dfrac{1}{\sqrt{1+L}},0,0 \right),\nonumber\hspace{0.2cm}
n^i=\left(\dfrac{1}{2},\dfrac{-A}{2\sqrt{1+L}},0,0 \right),\nonumber\\
&m^i=\left(0,0,\dfrac{1}{\sqrt{2}r},\dfrac{\rm icosec\theta}{\sqrt{2}r},\right),\cr
&\overline{m}^i=\left(0,0,\dfrac{1}{\sqrt{2}r},\dfrac{-\rm icosec\theta}{\sqrt{2}r}\right).
\end{align}

The twelve spin coefficients for the line element (1) are found as 
\begin{align}\label{D2}
&\kappa=\nu=\lambda=\sigma=\pi=\varepsilon=\tau=0,\nonumber\\
&\alpha=\dfrac{-\cot \theta}{2\sqrt{2}r},\rho=\dfrac{-1}{r\sqrt{1+L}},\hspace{0.4cm}
\mu=\dfrac{-A}{2r\sqrt{1+L}},\nonumber\\
&\beta=\dfrac{\cot \theta}{2\sqrt{2}r},\hspace{0.4cm}\gamma=\dfrac{A'}{4\sqrt{1+L}}.\hspace{0.4cm}
\end{align}
Taking the spinor of the form $G=G(r,\theta)e^{i(\omega t+\tilde{m}\phi)}$, where  $\tilde{m}$ and $\omega$ are the azimuthal quantum number  and  frequency of the wave corresponding to the Dirac particle respectively, we can solve the Chandrasekhar-Dirac equations. To obtain the angular parts and radial parts  from Eq. \eqref{D1}, the following transformations are taken as
\begin{align}\label{D3}
F_1= R_1(r)B_1(\theta)e^{i(\omega t+\tilde{m}\phi)},\nonumber\\
G_1= R_2(r)B_1(\theta)e^{i(\omega t+\tilde{m}\phi)},\nonumber\\
F_2= R_2(r)B_2(\theta)e^{i(\omega t+\tilde{m}\phi)},\nonumber\\
G_2= R_1(r)B_2(\theta)e^{i(\omega t+\tilde{m}\phi)}.
\end{align}
Using Eqs. \eqref{D2} and \eqref{D3} in  Eq. \eqref{D1}, we get
\begin{align}\label{D4}
&B_1\left[\dfrac{1}{\sqrt{1+L}}+\dfrac{ir\omega}{A}+\frac{i q Q}{A}+\dfrac{r}{\sqrt{1+L}}\dfrac{\partial}{\partial{r}}\right]R_1 \nonumber
\\
&+\dfrac{R_2}{\sqrt{2}}\widetilde L^+B_2=i\mu^*r B_1R_2,\nonumber\\
&AB_2\Big[\dfrac{1}{\sqrt{1+L}}+\dfrac{A'r}{2A\sqrt{1+L}}-\dfrac{ir\omega}{A}-\frac{i q Q}{A} \nonumber\\
&+\dfrac{r}{\sqrt{1+L}}\dfrac{\partial}{\partial{r}}\Big]R_2-\sqrt{2}R_1\widetilde LB_1=-2i\mu^*rB_2R_1,\nonumber\\
&B_2\left[\dfrac{1}{\sqrt{1+L}}+\frac{i q Q}{A}+\dfrac{ir\omega}{A}+\dfrac{r}{\sqrt{1+L}}\dfrac{\partial}{\partial{r}}\right]R_1 \nonumber\\
&-\dfrac{R_2}{\sqrt{2}}\widetilde LB_1=i\mu^*rB_2R_2,\nonumber\\
&AB_1\Big[\dfrac{1}{\sqrt{1+L}}+\dfrac{A'r}{2A\sqrt{1+L}}-\frac{i q Q}{A}-\dfrac{irw}{A}\nonumber\\
&+\dfrac{r}{\sqrt{1+L}}\dfrac{\partial}{\partial{r}}\Big]R_2+\sqrt{2}R_1\widetilde L^+B_2=-2i\mu^*rB_1R_1,\nonumber\\
\end{align}
where the angular operators are given by
\begin{align}\label{D5}
&\widetilde L^+=\dfrac{\partial}{\partial\theta}+\dfrac{m}{\sin\theta}+\dfrac{\cot\theta}{2},\cr
&\widetilde L^=\dfrac{\partial}{\partial\theta}-\dfrac{m}{\sin\theta}+\dfrac{\cot\theta}{2}.
\end{align}
Solving Eqs. \eqref{D4} and \eqref{D5}, the radial and angular parts can be 
expressed as 
\begin{align}
&\left[\dfrac{1}{\sqrt{1+L}}+\dfrac{i(r\omega+qQ)}{A}+\dfrac{r}{\sqrt{1+L}}\dfrac{\partial}{\partial{r}}\right]R_1-i\mu^*rR_2 \cr
&=\lambda_1R_2,\nonumber\\
&A\left[\dfrac{1}{\sqrt{1+L}}+\dfrac{A'r}{2A\sqrt{1+L}}-\dfrac{i(r\omega+qQ)}{A}+\dfrac{r}{\sqrt{1+L}}\dfrac{\partial}{\partial{r}} \right]R_2 \cr
&+2i\mu^*rR_1=\lambda_2R_1,\nonumber\\
&\left[\dfrac{1}{\sqrt{1+L}}+\dfrac{i(r\omega+qQ)}{A}+\dfrac{r}{\sqrt{1+L}}\dfrac{\partial}{\partial{r}} \right]R_1-i\mu^*rR_2 \cr
&=\lambda_3R_2,\nonumber\\
&A\left[\dfrac{1}{\sqrt{1+L}}+\dfrac{A'r}{2A\sqrt{1+L}}-\dfrac{i(r\omega+qQ)}{A}+\dfrac{r}{\sqrt{1+L}}\dfrac{\partial}{\partial{r}} \right]R_2 \cr
&+2i\mu^*rR_1=-\lambda_4R_1,
\end{align}
and
\begin{align}
\widetilde L^+B_2= -\sqrt{2}B_1\lambda_1,\hspace{1.5cm}\widetilde LB_1 =\dfrac{1}{\sqrt{2}}B_2\lambda_2,\nonumber\\
\widetilde LB_1 = \sqrt{2}B_2\lambda_3,\hspace{1.7cm}\widetilde L^+B_2 =-\dfrac{1}{\sqrt{2}}B_1\lambda_4,
\end{align}
where $\lambda_1,\lambda_2,\lambda_3$ and $\lambda_4$ denote the constants of separation. To derive the radial and angular parts from the Dirac equations, the constants of separation are defined by
$\lambda_1=\dfrac{\lambda_2}{2}=\dfrac{\lambda_3}{2}=\dfrac{\lambda_4}{4}=\lambda$. Taking $R_1=\dfrac{\psi_1 }{r}$ and $R_2=\dfrac{\psi_2}{r}$, we have\\
\begin{align}
&\dfrac{1}{\sqrt{1+L}}\left[\dfrac{d}{dr}+\dfrac{i\omega\sqrt{1+L}}{A}+\frac{i q Q\sqrt{1+L}}{r A}\right]\psi_1 \cr \label{rad1}
&=\left(\dfrac{\lambda}{r}+i\mu_*\right)\psi_2,\\ 
&\dfrac{A}{\sqrt{1+L}}\left[\dfrac{d}{dr}-\dfrac{i\omega\sqrt{1+L}}{A}-\frac{i q Q\sqrt{1+L}}{r A}+\dfrac{A'}{2A}\right]\psi_2 \cr \label{rad2}
&=\left(\dfrac{\lambda}{r}-i\mu_*\right)\psi_1,\\ 
&\tilde{L}^+B_2=-\lambda B_1 ,\hspace{1cm}
\tilde{L}B_1=\lambda B_2.
\end{align}
Applying tortoise coordinate defined by
\begin{align}\label{tortoise}
  dr_*=\dfrac{\Omega\sqrt{1+L}}{A}dr ,
\end{align} where $\Omega=1+\frac{qQ}{r\omega}$ and further taking the transformation\\                        $\psi=\tilde{R_1}(r)$, $\Psi_2=A^{-\frac{1}{2}}\tilde{R_2}(r)$,  the radial parts of Eqs. \eqref{rad1} and \eqref{rad2} can be written as 
\begin{align}
\left(\dfrac{d}{dr_*}+i\omega\right)\widetilde R_1=\frac{\sqrt{A}}{\Omega} \label{rad3} \left(\dfrac{\lambda}{r}+i\mu_*\right)\widetilde R_2,\\
\left(\dfrac{d}{dr_*}-i\omega\right)\widetilde R_2=\frac{\sqrt{A}}{\Omega} \label{rad4} \left(\dfrac{\lambda}{r}-i\mu_*\right)\widetilde R_1.
\end{align}

For massless Dirac field one can set $\mu_*=0$ in Eqs. \eqref{rad3} and \eqref{rad4}, the corresponding radial parts of Dirac equation become
\begin{align}\label{rad5}
\left(\dfrac{d}{dr_*}+i\omega\right)\tilde{R_1}=\frac{\sqrt{A}}{\Omega}\left(\dfrac{\lambda}{r}\right)\widetilde R_2,
\end{align}
\begin{align}\label{rad6}
\left(\dfrac{d}{dr_*}-i\omega\right)\tilde{R_2}=\frac{\sqrt{A}}{\Omega}\left(\dfrac{\lambda}{r}\right)\widetilde R_1.
\end{align}
Again, taking the transformation $Z_+ = \tilde{R_1}+\tilde{R_2}$ and $Z_-=\tilde{R_1}-\tilde{R_2}$, the Eqs. \eqref{rad5} and \eqref{rad6} reduce to
\begin{align}\label{rad7}
\left(\dfrac{d}{dr_*}-\frac{\sqrt{A}}{\Omega}\dfrac{\lambda}{r}\right)Z_+ =i\omega Z_- 
\end{align}
\begin{align}\label{rad8}
\text{and} \left(\dfrac{d}{dr_*}+\frac{\sqrt{A}}{\Omega}\dfrac{\lambda}{r}\right)Z_- =i\omega Z_+.
\end{align}
From Eqs. \eqref{rad7} and \eqref{rad8}, the pair of one dimensional wave equation resembling the Schrödinger equation   can be calculated as follows
\begin{align}\label{Waveeqn}
\left(\dfrac{d^2}{dr_*^2}+\omega^2\right)Z_{\pm}=V_{\pm}Z_{\pm},
\end{align}
where $V_+$ and $V_-$ denote the effective potentials for Dirac particles, which are given by 
\begin{align} \label{Veff}
V_\pm=\dfrac{A\lambda^2}{\Omega^2 r^2}\pm \frac{\sqrt{A} A^{'} \lambda}{2r\sqrt{1+L}\Omega^2}\mp \frac{A^{\frac{3}{2}}\lambda}{r^2 \Omega^3\sqrt{1+L}}.  
\end{align}

\begin{figure*}
\centering
\subfloat[]
{\includegraphics[width=175pt,height=155pt]{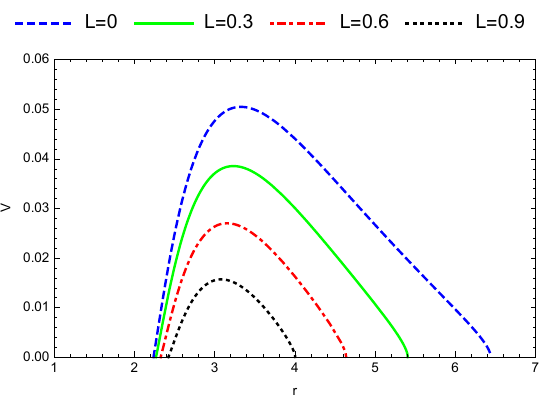}
\label {VmasslessL}
}
\hfill
\subfloat[]
{\includegraphics[width=175pt,height=155pt]{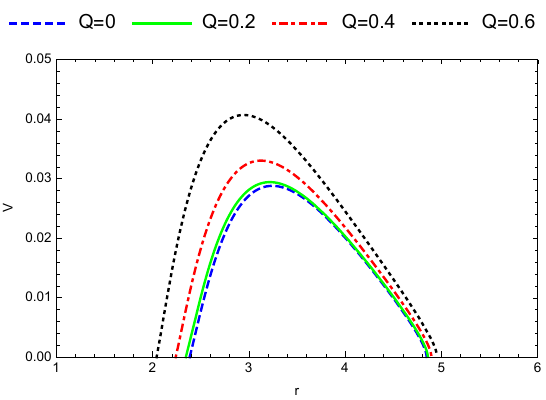}
\label {VmasslessQ}
}
\caption{Effective potential for massless Dirac perturbation with fixed $\Lambda=0.05$, $q=0.1$, $\omega=0.5$, $\ell=1$, $M=1$. (a) Variation in $L$ for $Q=0.3$. (b) Variation in $Q$ for $L=0.5$.}
\label{Vmassless}
\end{figure*}

It is found from Eq. \eqref{Veff} that the effective potential $V(r)$ of Dirac field depends on the BH parameters $Q$, $M$, $\ell$, $\Lambda$, $L$ and $q$. If $L=0$ in Eq. \eqref{Veff}, it becomes the effective potential of RNdS BH. Eq. \eqref{Veff} becomes the  effective potential of Schwarzschild-de Sitter-like BH if $Q=0$. We investigate the behaviour of effective potential of Dirac field as a function of the  radial coordinate $r$ for different values of $L$ and $Q$, as displayed in Figs. \ref{VmasslessL} and \ref{VmasslessQ}. It is evident that the effective potential $V(r)$ is positive definite between the event horizon and cosmological horizon and has a single maxima. In Fig. \ref{VmasslessL}, the peak of the effective potential decreases for large value of $L$ and the peak's position shifts toward the left. It is also found that the peak of the effective potential of RNdS BH is higher than the peak of the effective potential of RNdS-like BH. In Fig. \ref{VmasslessQ}, we observe that the peak of the effective potential increases with increasing $Q$ and the position of the peak moves toward the left. We also see that the peak of the effective potential of Schwarzschild-de Sitter-like BH is lower than the effective potential of RNdS-like BH. Hence, the presence of $L$ prevents the raise of the effective potential of RNdS-like BH but it has the opposite effect for the charge $Q$.

\section{Quasinormal modes}
By taking the corresponding conservation relation of the concerned spacetime and tortoise coordinate transformation, we derive the Schrodinger-like wave equation to investigate the QNMs. The QNMs of Dirac field perturbation are the solution of wave equation \eqref{Waveeqn} which holds special boundary conditions near the event horizon and far away from the BH spacetime \cite{Konoplya2011}. Its solution will obey the conditions of purely ingoing waves near the event horizon and purely outgoing wave at spatial infinity. 

\subsection{WKB Method}
In this subsection, the QNM frequencies will be studied by using Pade averaged sixth order WKB approximation. The QNM frequencies are complex numbers which describe the oscillation frequencies of perturbed BHs where the real part corresponds to the oscillation frequency and the imaginary part represents the damping rate. The WKB approximation was initially introduced by Schutz and Will \cite{Schutz1985} for calculating the QNM frequencies. Refs. \cite{Iyer1987a,Iyer1987b} improved the method by developing a third order WKB approximation. Further, Konoplya \cite{Konoplya2003} generalized the approach to higher orders. The formula for finding the QNM frequencies using sixth order WKB approximation reads as
\begin{align}\label{WKBquasi}
\frac{i(\omega^2-V_0)}{\sqrt{-2V_0^{''}}}-\sum\limits^6_{i=2} \Lambda_i=n+\frac{1}{2}.
\end{align}
Here, $n$ represents the overtone number and the subscript $0$ denotes the value of the variable at the point $r_*(r_0)$ at which the effective potential attains its maximum. $V_0^{''}$ is the second derivative of the effective potential with respect to tortoise coordinate at $r_0$. The terms $\Lambda_i (i=2,3,4,5,6)$ are presented in \cite{Iyer1987b, Konoplya2003}. An accurate result of WKB method is obtained only when the multipole number ($\ell$) is bigger than the overtone number $(n)$ and it gives less accurate when $n>\ell$. To obtain better results of QNMs of higher order, Ref. \cite{matyjasek2017,matyjasek2019} developed Pade approximation based on WKB method.

\subsection{AIM Method}
In this subsection, we will calculate QNMs by using the Asymptotic Iteration
Method (AIM). AIM  is a numerical method developed for solving homogeneous second-order ordinary differential equations \cite{Ciftci}. Due to the accuracy and computing efficiency, the AIM has been widely used in many different areas of physics including quantum mechanics,  BH perturbation theory.  Using AIM method, one can accurately calculate the QNM frequencies by expanding the solution of the perturbation equation around a regular point and applying a recursive relation between successive derivatives. It was applied to calculate the QNM frequencies of field perturbation in Schwarzschild BH spacetime in asymptotically flat and dS spacetimes \cite{Cho2010}. Subsequently this method is widely used to investigate the QNM frequencies for different types of BHs in different modified theories of gravity \cite{Pong2018,Pong2019,Pong2020,Pong2022,Liu2023}. In this study, we will use the AIM to solve the radial wave equation \eqref{Waveeqn} and compute the QNM frequencies of the Dirac field perturbation in the RNdS-like BH for varying $Q$ and $L$. Introducing a new variable $u=1/r$,  Eq. \eqref{Waveeqn} reduces to
\begin{align}
&(\Omega^2 - V(u)) \Psi(u) + \frac{p(u) p'(u) \Psi'(u)}{(1 + L) \Omega(u)^2} - \frac{p(u)^2 \Psi'(u) \Omega'(u)}{(1 + L) \Omega(u)^3}\cr
& + \frac{p(u)^2 \Psi''(u)}{(1 + L) \Omega(u)^2}=0,
\end{align}
where
\begin{align}
p(u) = u^2 - 2 M u^3 + \frac{2 (1 + L) Q^2 u^4}{2+L}-\frac{(1+L)\Lambda}{3}. 
\end{align}
The metric function $f(u)$ can be written in the form
\begin{align}
f(u)=\dfrac{1}{u^2} (u-u_1) (u-u_2)(u-u_3)(u-u_4).
\end{align}
The surface gravity is calculated as
\begin{align}
\kappa_i&=\dfrac{1}{2} \left.\dfrac{df(r)}{dr}\right\vert_{r\rightarrow r_i}=- \dfrac{u^2}{2} \left.\dfrac{df(u)}{du}\right\vert_{u\rightarrow u_i}\cr
&=-\dfrac{1}{2} \prod_{\substack{i \ne j}} (u_i-u_j).
\end{align}
The tortoise coordinate  \eqref{tortoise} is rewritten using the new variable $u$ as
\begin{align}\label{tortoise2}
r_*=&-\int \dfrac{\sqrt{1+L} (1+u \Xi)}{u^2 f(u)} \nonumber\\
=& -\int \sum_{i=1}^4 \dfrac{A_i}{(u-u_i)} du,
\end{align}
where $\Xi=q\,Q/\omega$. From Eq. \eqref{tortoise2}, we can obtain the expression of $A_i$ by calculating
\begin{align}\label{tortoise3}
\sqrt{1+L} (1+u \Xi)= \sum_{i=1}^4 \left(A_i \prod_{\substack{j \ne i}} (u-u_j)\right).
\end{align}
On solving Eq. \eqref{tortoise3}, we obtain
\begin{align}\label{tortoise4}
A_i=\dfrac{-\sqrt{1+L} (1+u_i \Xi)}{2 \kappa_i}.
\end{align}
Thus the tortoise coordinate $r_*$ takes the form
\begin{align}\label{tortoise5}
r_*=\ln \left( \prod_{\substack{i =1}}^4 (u-u_i)^	{\dfrac{\sqrt{1+L} (1+u_i \Xi)}{2 \kappa_i}}\right).
\end{align}
To scale out the divergent behavior  at the cosmological horizon, the wave function is taken as
\begin{align}\label{tortoise6}
Z=e^{i \omega r_*} \xi(u).
\end{align}
Further to scale out the divergent behavior at the event horizon, we choose the function $\xi(u)$ as 
\begin{align}\label{tortoise7}
\xi(u)=(u-u_1)^{-i\omega/2\kappa_1} \chi(u).
\end{align}
Here $u_1=1/r_1$, where $r_1$ is the event horizon of the BH and $\kappa_1$ is the surface gravity at $r_1$. Using Eqs. \eqref{tortoise6} and \eqref{tortoise7}, the wave equation \eqref{tortoise2} reduces to
 \begin{align}\label{chi}
 \chi''(u)=\lambda_0(u) \chi'(u)+s_0(u) \chi(u),
\end{align}  
where
\begin{align}
\lambda_0&=\frac{2i \sqrt{1+L}\omega \Omega(u_1)}{(u - u_1)\kappa(u_1)}+\frac{2i \sqrt{1+L}\omega \Omega(u)-p'(u)}{p(u)} \nonumber \\ &+\frac{\Omega'(u)}{\Omega(u)},\cr
s_0&=\frac{(1 + L) \omega^2 \Omega(u_1)^2}{(u - u_1)\kappa(u_1)^2}+\frac{\sqrt{1+L} \lambda}{6 (2 + L) \omega1 p(u)^{\frac{3}{2}} \Omega(u)}\cr
&\times \Big[2 q Q (-3 (2 + L) M u^3 + (1 + L) 6 Q^2 u^4\cr
& + (2 + L) \Lambda))+6\sqrt{1+L}(2+L)\lambda \omega1 \sqrt{p(u)} \Omega(u)\cr
&+3(2+L) \omega1 p'(u)\Big]\cr
&+\frac{\sqrt{1+L}\omega \Omega(u1)W}{(u-u_1)^2 p(u) \kappa(u_1) \Omega(u_1)},
\end{align}
where 
\begin{align*}
W&=(u-u_1)\Omega(u)(2\sqrt{1+L}\omega\Omega(u)+i p'(u)) \cr
&-ip(u) \left\lbrace \Omega(u)+(u-u_1)\Omega'(u)\right\rbrace.
\end{align*}
Now taking n$^{th}$ derivative of Eq. \eqref{chi}, one can readily obtain \cite{Cho2010}
 \begin{align}
\chi^{(n+2)}=\lambda_n(u) \chi'(u)+s_n(u) \chi(u),
\end{align}
where
\begin{align}
& s_n=s'_{n-1}+s_0 \lambda_{n-1}\label{sn},\\
&\lambda_n=\lambda'_{n-1}+\lambda_{n-1} \lambda_0+s_{n-1}. \label{lambdan}
\end{align}
We will use the improved version of AIM  proposed in \cite{Cho2010}. For this we expand $\lambda_n$ and $s_n$ around some arbitrary point $\tilde{u}$ in the Taylor series as follows \cite{Cho2010}
\begin{align}
& \lambda_n(u)=\sum_{i=0}^\infty c_{n}^i (u-\tilde{u})^i, \cr
& s_n(u)=\sum_{i=0}^\infty d_{n}^i (u-\tilde{u})^i,
\end{align}
where $c_{n}^i$ and $d_{n}^i$ represent the $i$th coefficients in the Taylor series expansions  of $\lambda_n$ and $s_n$ respectively. Using these expressions, Eqs. \eqref{lambdan} and \eqref{sn} can be written as
\begin{align}
&c_{n}^i=(i+1) c_{n-1}^{i+1}+d_{n-1}^i+\sum_{k=0}^i c_{0}^k c_{n-1}^{i-k},\\
& d_{n}^i=(i+1) d_{n-1}^{i+1}+\sum_{k=0}^i d_{0}^k c_{n-1}^{i-k}.
\end{align}
Now the quantization condition is expressed as
\begin{align}\label{AIM}
d_{n}^0 c_{n-1}^0-d_{n-1}^0 c_{n}^0=0.
\end{align}
The QNM frequencies are evaluated by solving the above recursion relation \eqref{AIM}. It is important to note that the improved AIM depends on the choice of an expansion point $\tilde{u}$. The fastest convergence of AIM is observed when $\tilde{u}$ is taken as the location of the maximum of the effective potential \cite{Barakat}. \\
The numerical values of QNM frequencies using sixth-order WKB method and AIM method with varying $L$ and $Q$ for the RNdS-like BH are displayed in Tables \ref{QNML} and \ref{QNMQ} respectively. For fixed $\ell, M, \Lambda, Q, q$ and $\omega$, increasing the value of $L$, both the real part and the absolute value of the imaginary part of QNM frequencies decrease for the two methods but it has the opposite effect for the increase of $Q$. We also evaluate the root mean square (rms) error associated with the WKB and AIM methods.

\begin{sideways}
\begin{minipage}{\textheight}
\centering
\captionof{table}{The QNMs frequencies for varying $L$ with fixed $\Lambda = 0.05, M = 1, Q = 0.2$ and $q = 0.1$  using WKB and AIM methods.}\label{QNML}
\begin{tabular}{c|ccc|ccc}
 & \multicolumn{3}{c|}{WKB 6th order} & \multicolumn{3}{c}{AIM}  \\ \hline
 $L$& $\ell=2$ & $\ell=3$           &  $\ell=5$        & $\ell=2$ &   $\ell=3$       &    $\ell=5$   \\
0 & 0.354837-0.0707172i  &  0.497842 - 0.0707498i & 0.783368 - 0.0707699i & 0.354275-0.0704672i & 0.497834 -0.0709088i &   0.783364 -0.0707616i \\
 0.2 & 0.326134-0.0592572i & 0.457272 - 0.0592821i  &   0.719239 - 0.0592976i  & 0.325801-0.0592999i & 0.457317 -0.0593507i  &   0.719235 -0.0592950i    \\
 0.4& 0.294279-0.0494613i &  0.412417 - 0.0494809i   &  0.648502 - 0.0494927i  & 0.29417-0.0495586i &   0.412456 -0.0494980i  &   0.648500 -0.0494929i     \\
 0.6& 0.258253-0.0405845i & 0.361804 - 0.0405973i &   0.568796 - 0.0406049i   & 0.258303-0.0405765i &  0.361825 -0.0405939i    &    0.568795 -0.0406055i   \\
 0.8 & 0.216081-0.0320079i    &  0.302643 - 0.0320149i   & 0.475709 - 0.0320191i             & 0.216174-0.0316989i &  0.302649 -0.0320090i  &  0.475709 -0.0320195i  
\end{tabular}
\end{minipage}
\end{sideways}

\begin{sideways}
\begin{minipage}{\textheight}
\centering
\captionof{table}{The QNMs frequencies for varying $Q$ with fixed $L= 0.6,  M = 1, \Lambda = 0.05$,  and $ q = 0.1$ using WKB and AIM methods.}\label{QNMQ}
\begin{tabular}{c|ccc|ccc}
 & \multicolumn{3}{c|}{WKB 6th order} & \multicolumn{3}{c}{AIM}  \\ \hline
 $Q$&  $\ell=2$ & $\ell=3$           &  $\ell=5$        &   $\ell=2$& $\ell=3$       &    $\ell=5$     \\
0 & 0.254227-0.0402290i & 0.356164 - 0.0402415i    &   0.559928 - 0.0402488i       & 0.25426-0.0402373i & 0.356186 -0.0402394i  &    0.559927 -0.0402495i  \\
 0.2& 0.258253-0.0405845i &  0.361804 - 0.0405973i    &   0.568796 - 0.0406049i       & 0.258303-0.0405765i & 0.361825 -0.0405939i  &     0.568795 -0.0406055i   \\
 0.4& 0.277305-0.0426918i &  0.38851 - 0.042706i    &  0.610793 - 0.0427144i         & 0.277334-0.0427069i & 0.388528 -0.0426998i   &    0.610792 -0.0427151i   \\
 0.6& 0.313771-0.0462235i  & 0.439621 - 0.0462414i    &   0.691167 - 0.0462511i       & 0.313793-0.046253i &  0.439632 -0.0462312i  &         0.691167 -0.0462517i \\
 0.8&  0.377706-0.04990489i &  0.529235 - 0.0499323i  &     0.832092 - 0.0499476i     & 0.377605-0.0498387i  &  0.529233 -0.0499311i  &  0.832092 -0.0499480i      
\end{tabular}
\end{minipage}
\end{sideways}

\begin{table*}[]
\caption{Numerical calculation of root mean square error between WKB and AIM methods for varying $Q$ and $L$.}\label{RMS}
\begin{tabular}{c|ccc|c|ccc}
  \multicolumn{4}{c}{$\Delta_{rms}$} & \multicolumn{4}{|c}{$\Delta_{rms}$}  \\ \hline
  $L$&  $\ell=2$ & $\ell=3$ &  $\ell=5$ & $Q$ &  $\ell=2$ & $\ell=3$ & $\ell=5$   \\
 0 &  6.15097$\times 10^{-4}$ &   1.59201 $\times 10^{-4}$ & 9.21358 $\times 10^{-6}$    & 0 & 3.40278 $\times 10^{-5}$ & 2.21 $\times 10^{-5}$  &  1.22066 $\times 10^{-6}$  \\
 0.2 & 3.35727  $\times 10^{-4}$& 8.20424 $\times 10^{-5}$ &  4.77074 $\times 10^{-6}$  & 0.2 & 5.0636  $\times 10^{-5}$&  2.12735 $\times 10^{-5}$ & 1.16619 $\times 10^{-6}$  \\
 0.4 & 1.46111 $\times 10^{-4}$ &   4.25842 $\times 10^{-5}$ & 2.00998 $\times 10^{-6}$   & 0.4 & 3.26957 $\times 10^{-5}$ &   1.90379 $\times 10^{-5}$ & 1.22066 $\times 10^{-6}$  \\
 0.6 & 5.0636 $\times 10^{-5}$ & 2.12735 $\times 10^{-5}$ & 1.16619 $\times 10^{-6}$   & 0.6 & 3.68001  $\times 10^{-5}$&   1.50013 $\times 10^{-5}$ &  6.$\times 10^{-7}$  \\
 0.8 & 3.22692 $\times 10^{-4}$ &  8.41487 $\times 10^{-6}$ &   4.$\times 10^{-7}$ & 0.8 & 1.20762  $\times 10^{-4}$&   2.33238 $\times 10^{-6}$   &  4.$\times 10^{-7}$  
\end{tabular}
\end{table*}

\section{Detectability of Quasinormal Modes with Gravitational Wave Detectors}
This section briefly explores the potential for detecting QNMs of RNdS-like BH in bumblebee gravity. It is necessary to convert the QNM frequencies into physical units to assess their potential detectabilities through gravitational wave observations.  Using Ref. \cite{Ferrari2008}, we assume the BH has a mass $M=\hat{\eta} M_\odot$, where $\hat{\eta}$ is a dimensionless scaling parameter and $M_\odot=1.48 \times 10^5$cm, then the real and the imaginary parts of the QNM frequencies are  transformed to a physical oscillation frequency and a decay timescale as
\begin{align}
& \hat{f}=\dfrac{c \,M \,\omega_R}{2\pi \hat{\eta} \,M_\odot } \,\text{kHz}=\dfrac{32.36 \times M \, \omega_R}{\hat{\eta}} \, \text{kHz}, \nonumber\\
\text{and} \quad& \tau=\dfrac{\hat{\eta} \,M_\odot}{M \,\omega_i \, c}\,s= \dfrac{\hat{\eta} \times 0.4937 \times 10^{-5}}{M \omega_i} \,s.
\end{align}
Using these expressions one can assess whether the gravitational wave signals associated  with the Dirac perturbation in bumblebee gravity  are potentially detectable by the current and future gravitational wave observatories. 
Ground-based gravitational wave detectors, such as LIGO and Virgo can detect gravitational waves within the frequency range of approximately 10 Hz to 1000 Hz. In contrast, space-based detectors like LISA are 
sensitive to much lower frequencies, typically in the range of $10^{-4}$ Hz to 1 Hz. 
We compute the corresponding minimum and maximum BH masses in which the Dirac quasinormal modes fall within the sensitivity bands of LIGO and LISA. For this analysis,  we use the QNM frequency for $\ell=2$ calculated using  the Padé averaged sixth-order WKB method.

\begin{table*}[h!]
\centering
\begin{tabular}{p{1.5cm} p{1.8cm} p{1.8cm} p{1.8cm} p{2.5cm} p{2cm}}
\hline
$L$ & $\omega_R$ & $M_{\min}^{\text{LIGO}}$ & $M_{\max}^{\text{LIGO}}$ & \,\,\, $M_{\min}^{\text{LISA}}$ & $M_{\max}^{\text{LISA}}$ \\
\hline
0   & 0.354837   & 9.5392     & 953.9201    & 1.14470416 $\times 10^4$  & $1.14 \times 10^8$ \\
0.2 & 0.326134   & 8.7676     & 876.7569    & 1.05210828 $\times 10^4$  & $1.05 \times 10^8$ \\
0.4 & 0.294279   & 7.9112     & 791.1200    & 9.4934405 $\times 10^3$   & $9.49 \times 10^7$ \\
0.6 & 0.258253   & 6.9427     & 694.2701    & 8.3312418  $\times 10^3$  & $8.33 \times 10^7$ \\
0.8 & 0.216081   & 5.8090     & 580.8978    & 6.9707731 $\times 10^3$   & $6.97 \times 10^7$ \\
\hline
\end{tabular}
\caption{Minimum and maximum detectable BH masses for LIGO and LISA as a function of  $L$.}
\label{tab LIGOL}
\end{table*}

\begin{table*}[h!]
\centering
\begin{tabular}{p{1.5cm} p{1.8cm} p{1.8cm} p{1.8cm} p{2.5cm} p{2cm}}
\hline
$Q$ & $\omega_R$ & $M_{\min}^{\text{LIGO}}$ & $M_{\max}^{\text{LIGO}}$ & \,\,\, $M_{\min}^{\text{LISA}}$ & $M_{\max}^{\text{LISA}}$ \\
\hline
0   & 0.254227   & 6.8345     & 683.4469    & 8.2013630 $\times 10^3$    & $8.20 \times 10^7$ \\
0.2 & 0.258253   & 6.9427     & 694.2701    & 8.3312418 $\times 10^3$    & $8.33 \times 10^7$ \\
0.4 & 0.277305   & 7.4549     & 745.4883    & 8.9458593 $\times 10^3$    & $8.95 \times 10^7$ \\
0.6 & 0.313771   & 8.4352     & 843.5210    & 1.01222525 $\times 10^4$   & $1.01 \times 10^8$ \\
0.8 & 0.377706   & 10.1540    & 1015.3996   & 1.21847956  $\times 10^4$  & $1.22 \times 10^8$ \\
\hline
\end{tabular}
\caption{Minimum and maximum detectable BH masses for LIGO and LISA as a function of the charge  $Q$.}
\label{tab LIGOQ}
\end{table*}

Tables \ref{tab LIGOL} and \ref{tab LIGOQ} summarize the detectability mass range for different values of $L$ and $Q$ respectively. As $L$ increases, the detectable BH mass range shifts to lower values for both LIGO and LISA detectors. In contrast, increasing the BH charge 
$Q$ extends the detectable mass range toward higher values, enabling the detection of more massive BHs within the sensitivity limits of LIGO and LISA.
It is evident from Tables \ref{tab LIGOL} and \ref{tab LIGOQ} that neither Sgr A* nor M87* falls within the detectable mass range of LIGO. Further the mass of Sgr A* lies  within LISA's detection band for all considered values of  $L$ and $Q$. However the mass of M87* exceeds the upper detection limits, indicating that its QNMs signals would fall outside LISA’s sensitivity band. Our analysis highlights how Lorentz-violating effects and BH charge alter the detectable mass range for gravitational waves, offering key insights that can support the interpretation of future observations.

\section{Greybody factor}

In this section, we compute the GF associated with the scattering of Dirac field perturbations for RNdS-like BH. The GF is defined as the probability of an outgoing wave successfully traverses the spacetime potential barrier and reaches a distant observer. In the context of BH
scattering processes the GF are connected to the transmission  coefficient. WKB method is widely used to calculate the transmission coefficients and reflection coefficients for different types of fields perturbations in different BHs. From Figs. \ref{VmasslessL} and \ref{VmasslessQ}, it is observed that the outer boundary  of the effective potential barrier is the cosmological horizon. Thus any wave propagating toward the cosmological horizon encounters this potential barrier. As a result some waves are reflected back toward the BH and some waves manage to pass the potential barrier. This scattering process can be represented by the wave function
\begin{align}
&Z(r_*) = 
T(\omega)\,e^{-i\omega r_*},  \quad \quad \quad \quad~ r_* \to -\infty, \\
& Z(r_*)= e^{-i\omega r_*} + R(\omega)\,e^{i\omega r_*},  ~~~ r_* \to +\infty,
\end{align}
where $T(\omega)$ and $R(\omega)$ represent the  transmission  and reflection coefficients respectively. Using the WKB approximation, the reflection coefficient can be expressed as
\begin{align}
R=\left(1+e^{-2 i \pi K}	\right)^{-\frac{1}{2}},
\end{align}
where $K$ can be calculated from the equation
\begin{align}
K-i \dfrac{\left(\omega^2-V_0 \right)}{\sqrt{-2 V_{0}''}}-\sum_{i=2}^{i=6} \Lambda_{i}(K)=0,
\end{align}
where $\Lambda_i$ denotes the same correction term as given in Eq. \eqref{WKBquasi}.
Based on the conservation of probability, the transmission coefficient and the reflection coefficient satisfy the relation
\begin{align}
\vert T \vert^2+ \vert R \vert^2=1.
\end{align}
The GF $\gamma_\ell$ is defined as the transmission coefficient is obtained as
\begin{align}
\gamma_\ell=\vert T \vert^2=\left(1+e^{2 i \pi K}	\right)^{-1}.
\end{align}
The above formula is widely used to calculate the GFs for various BHs and wormholes \cite{konoplya2019a,konoplya2019b,konoplya2020}. In the low-frequency regime, the accuracy of the WKB method becomes lower as reflection dominates and GFs tends to zero.

\begin{figure*}[!htbp]
\centering
\subfloat[]
{\includegraphics[width=175pt,height=155pt]{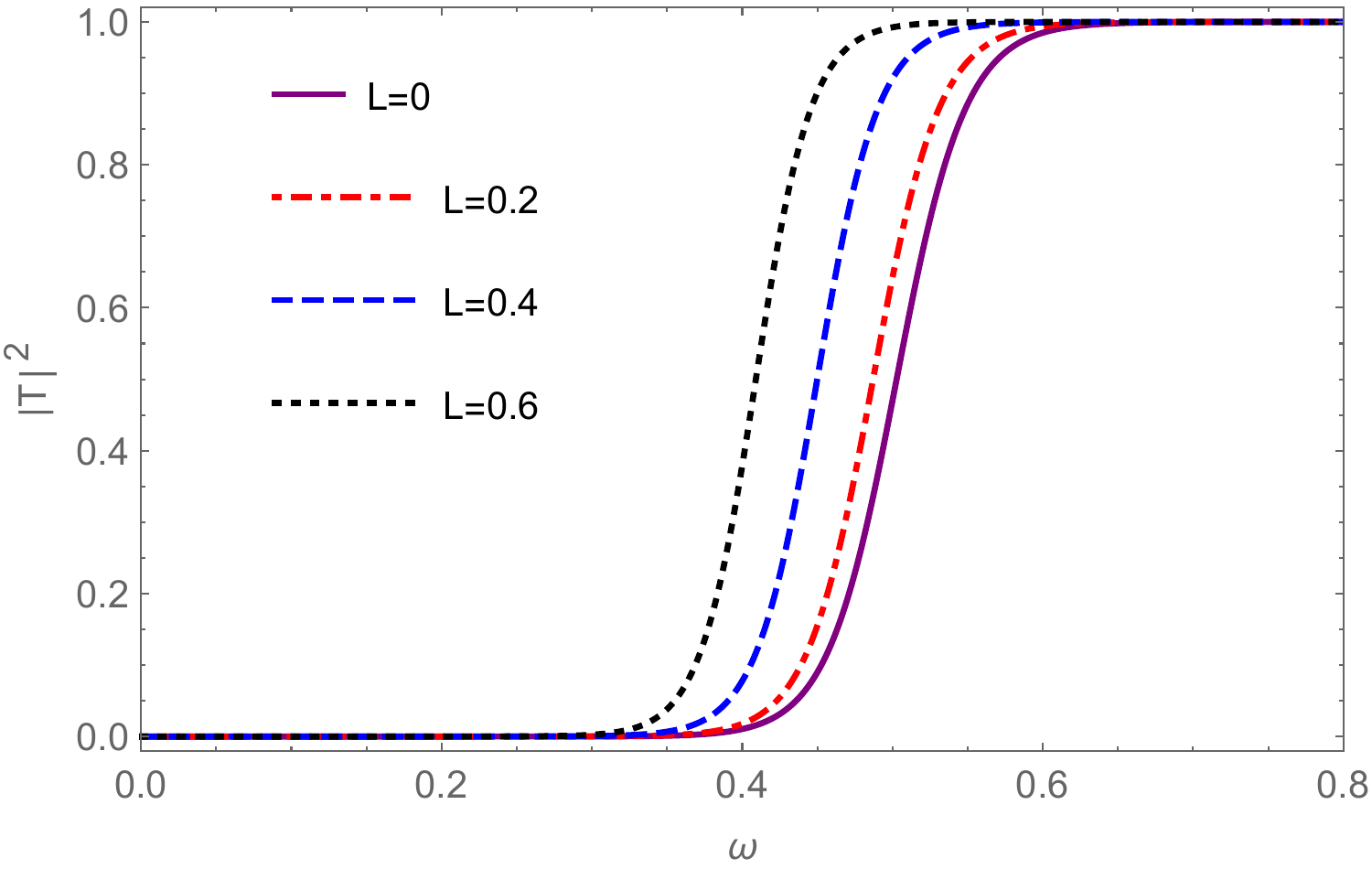}
\label {transmissionL}
}
\hfill
\subfloat[]
{\includegraphics[width=175pt,height=155pt]{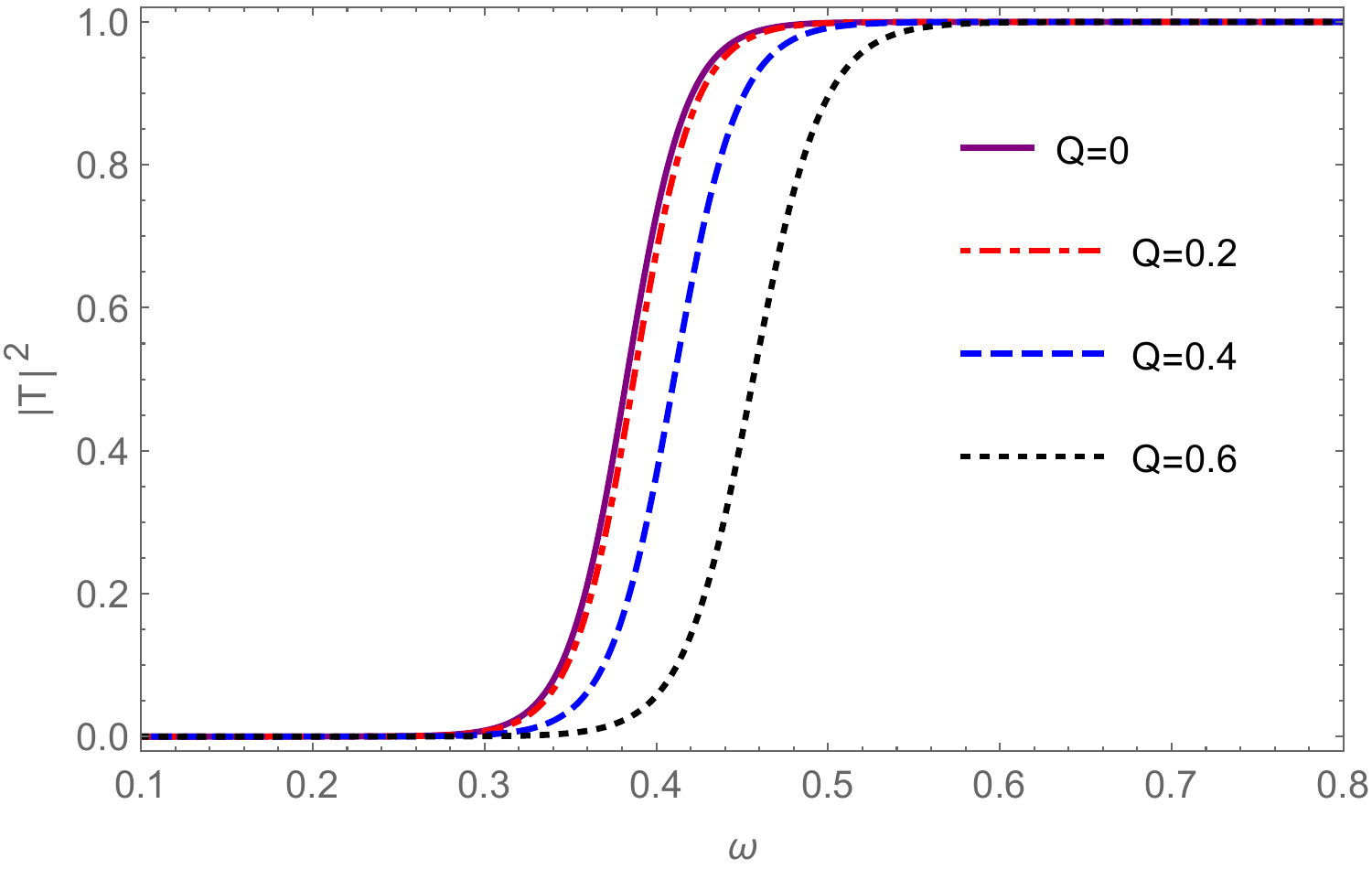}
\label {transmissionQ}
}
\caption{GF for massless Dirac perturbation with fixed $\Lambda=0.05$, $q=0.1$,  $\ell=3$, $M=1$. (a) Variation in $L$ for $Q=0.3$. (b) Variation in $Q$ for $L=0.5$. }
\label{transmission}
\end{figure*}

In Figs. \ref{transmissionL} and \ref{transmissionQ}, we illustrate the GF and examined the impact of $L$ and $Q$ respectively.  The case $L=0$ represents the GF of the standard RNdS BH. One can see that the GF of RNdS-like BH is higher than that of RNdS case and further it increases with increasing $L$. This suggests that the presence of Lorentz violation theory allows more wave to transmit. Moreover, the GF exhibits a decreasing trend with increasing $Q$ indicating that higher charge reduces the probability of wave transmission. The opposite effects of $L$ and $Q$ on the GF are consistent with the behaviour of effective potential obtained earlier in which  increasing $L$ reduces the height of the  effective potential barrier, thereby enhancing transmission, while increasing $Q$ raises the barrier, leading to greater reflection and reduces  the transmission.

\section{Absorption cross section}
Following the study of GF, we will study  the influence of Lorentz violation in bumblebee gravity on the absorption characteristics. The absorption cross section is a fundamental quantity that measures  the amount of the incoming field absorbed by the BH rather than scattering. It  provides a vital information about the interaction between perturbing fields and the BH geometry. According to quantum mechanics, the partial absorption cross section for each angular mode $\ell$ can be calculated using the transmission coefficient as
\begin{align}
\sigma_\ell= \dfrac{\pi (2\ell+1)}{\omega^2} \vert T_\ell \vert^2.
\end{align}
By taking the sum of all the partial absorption cross sections,  the total absorption cross section is given by
\begin{align}
\sigma= \dfrac{\pi}{\omega^2}\sum_{\ell} (2\ell+1) \vert T_\ell \vert^2.
\end{align}

\begin{figure*}[!htbp]
\centering
\subfloat[]
{\includegraphics[width=175pt,height=155pt]{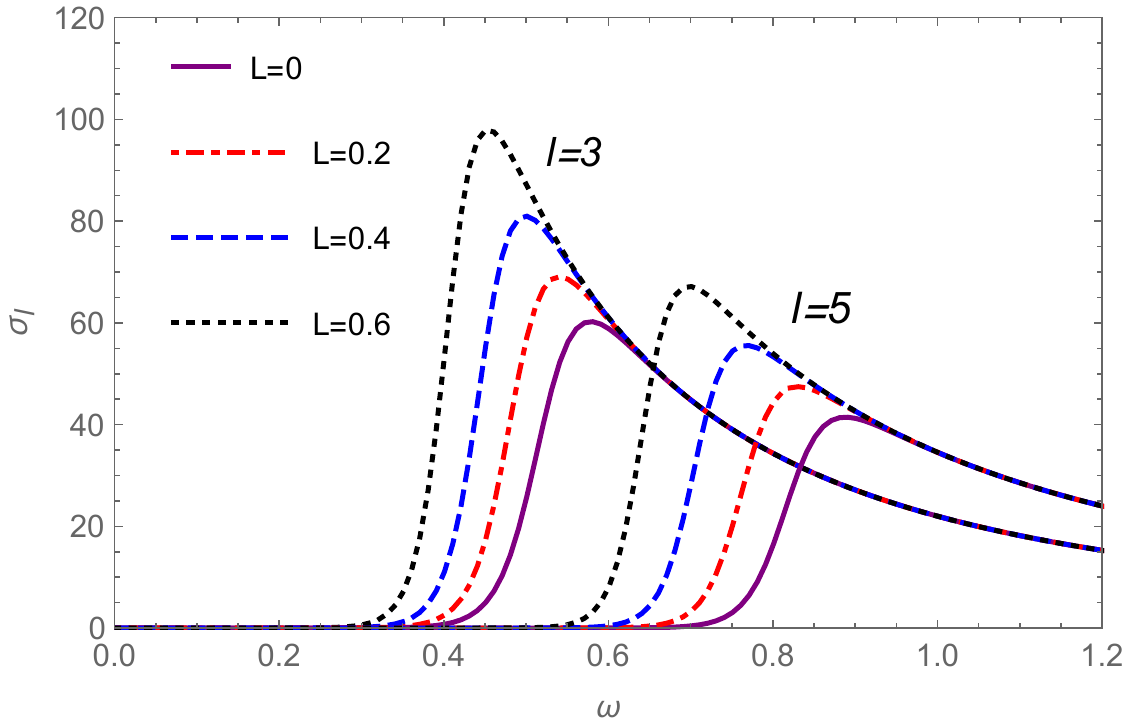}
\label {abspartialL}
}
\hfill
\subfloat[]
{\includegraphics[width=175pt,height=155pt]{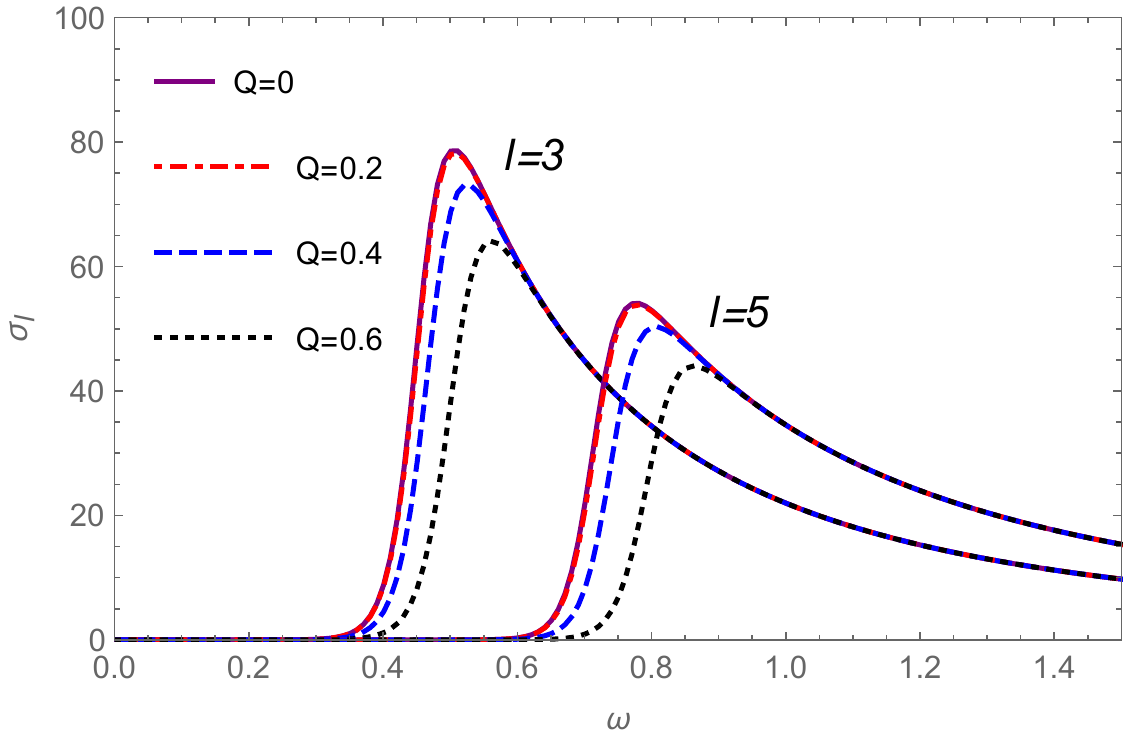}
\label {abspartialQ}
}
\caption{ Partial absorption cross section versus of $\omega$ for $l=3$ and 5 with fixed $\Lambda=0.05$, $q=0.1$,   $M=1$.  (a) Variation in $L$ for $Q=0.5$. (b) Variation in $Q$ for $L=0.2$.}
\label{abspartial}
\end{figure*}
\begin{figure*}[!htbp]
\centering
\subfloat[]
{\includegraphics[width=175pt,height=155pt]{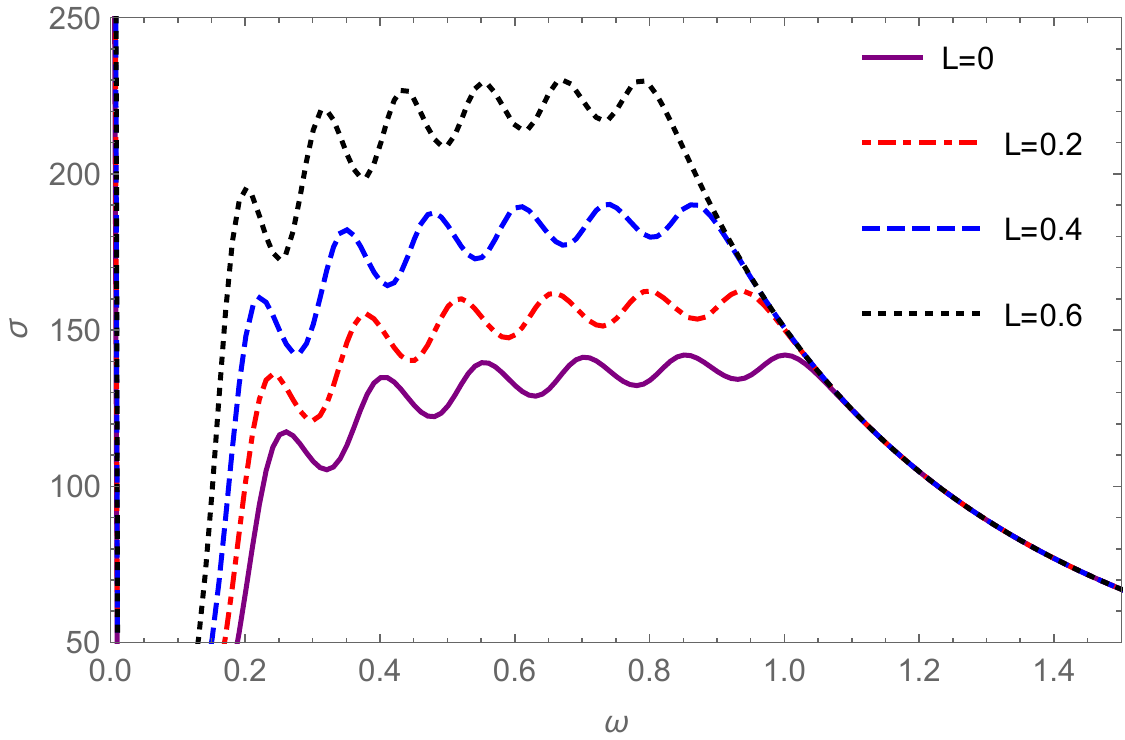}
\label {absL}
}
\hfill
\subfloat[]
{\includegraphics[width=175pt,height=155pt]{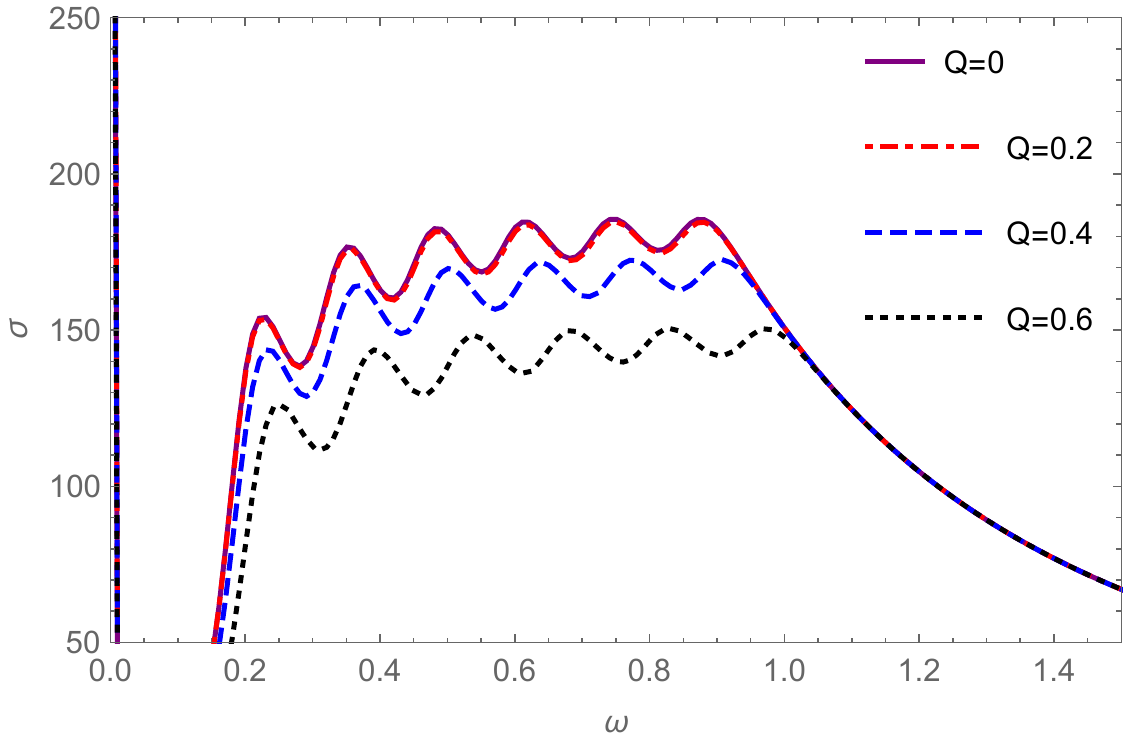}
\label {absQ}
}
\caption{ The total absorption cross section  versus  $\omega$  with fixed $\Lambda=0.05$, $q=0.1$,   $M=1$, calculated by summing over multipoles up to $\ell=6$  (a) Variation in $L$ for $Q=0.5$. (b) Variation in $Q$ for $L=0.2$. }
\label{totalabs}
\end{figure*}

The partial absorption cross sections corresponding to $\ell=$ 3  and $\ell=5$ are shown in Figs. \ref{abspartialL} and \ref{abspartialQ} respectively, to illustrate the effects of  $L$ and $Q$. The figure shows that increasing the values of $L$ results lower partial absorption cross sections while increasing $Q$ leads to higher partial absorption cross sections. This observation is in agreement with the characteristics  of the effective potential, wherein the  potential barrier becomes weaker with increasing $L$ resulting in higher absorption. In contrast, increasing $Q$ raises the potential barrier thereby  suppressing the transmission of the field  and results in lower absorption. In the high-frequency limit, the cross sections for different values of $L$ and $Q$ converge and thus the influence of $L$ and $Q$ become insignificant. The total absorption cross sections for different values of $L$ and $Q$ are plotted in Figs. \ref{absL} and \ref{absQ} respectively. In this particular example, we add the partial absorption cross section upto $\ell=6$. It is evident that the total absorption cross section in RNdS-like BH in bumblebee gravity ($L \ne 0$) is higher than the RNdS blak hole ($L=0$).

\section{Sparsity of Hawking emission}
In this section, we discuss the behavior of Hawking sparsity for massless Dirac field perturbation under the influence of L and charge $Q$. The Hawking temperature of RNdS-like BH is calculated as
\begin{align}
T_H&=\frac{1}{4\pi \sqrt{-g_{tt}g_{rr}}}\frac{dg_{tt}}{dr}|_{r=r_h} \cr
&=\frac{1}{2\pi \sqrt{1+L}}\left(\frac{M}{r_h^2}-\frac{r\Lambda(1+L)}{3}-\frac{2Q^2(1+L)}{r_h^3(2+L)}\right). \nonumber\\
\end{align}
For a BH  radiating at temperature $T_H$, the total power of Hawking radiation corresponding to the frequency $\omega$ in the momentum  interval $d^3k$,  is given by \cite{Gray,Miao}
\begin{eqnarray}\label{ps1}
\frac{d E(\omega)}{dt}\equiv P_{tot}=\sum_{\ell} \gamma_{\ell} \frac{\omega}{e^{\omega/T_0}-1} \hat{k}. \hat{n} \frac{d^3k\,\, d A}{(2\pi)^3},
\end{eqnarray}
where $\hat{n}$ and  $\gamma_{\ell}$ are the unit normal vector of the surface element $dA$  and the GF respectively. We consider $|k|=\omega$ for massless particle in Eq. \eqref{ps1}, the total power of Hawking radiation becomes 
\begin{eqnarray}\label{ps2}
P_{tot}=\sum_{\ell} \int_0^\infty P_{\ell}(\omega) d\omega.
\end{eqnarray}
Here, $P_{\ell}(\omega)$ represents the power spectrum in the $\ell^{th}$ mode which is given by
\begin{eqnarray}\label{ps3}
P_{\ell}(\omega)=\frac{A_h}{8\pi^2} \gamma_{\ell}\frac{\omega^3}{e^{\omega/T_0}-1},
\end{eqnarray}
where $A_h$ denotes a multiple of the horizon area of RNdS-like BH.\\
To gain more insight into the radiation emitted by BH, a dimensionless parameter $\eta$ known as the sparsity of Hawking radiation is introduced and is defined by \cite{Gray,Miao,Hod2015,Hod2016,Chow2020}
\begin{align}
\eta=\frac{\tau_{gap}}{\tau_{emission}},
\end{align}
where $\tau_{gap}$ and $\tau_{emission}$ are the average time interval between the emission of two successive Hawking radiation quanta and the characteristic time for the emission of individual Hawking quantum respectively. These quantities are also defined by
\begin{align}
\tau_{gap}=\frac{\omega_{max}}{P_{tot}}, \tau_{emission} \geq \tau_{localisation}=\frac{2\pi}{\omega_{max}}.
\end{align} 

Here $\tau_{localisation}$ represents the time taken by the emitted wave of frequency $\omega_{max}$ to complete one cycle of oscillation. Hawking radiation is continuous if $\eta\leqslant1$ and a sparse Hawking radiation occurs when $\eta$ is very large.\\

\begin{figure*}[!htbp]
\centering
\subfloat[]
{\includegraphics[width=175pt,height=155pt]{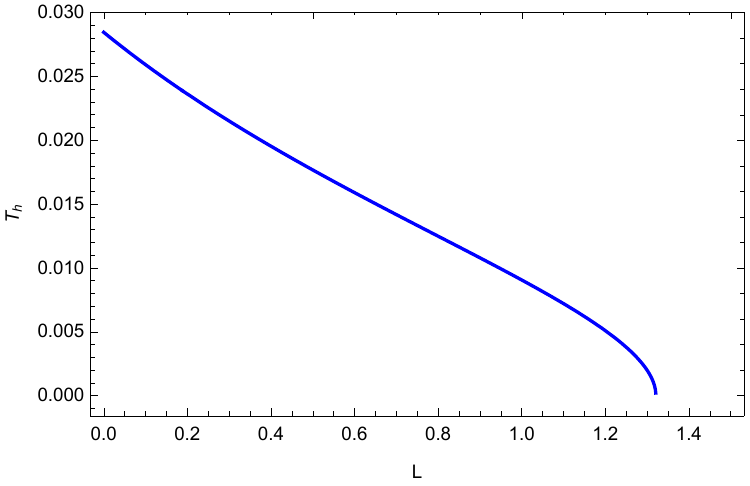}
\label {tempL}
}
\hfill
\subfloat[]
{\includegraphics[width=175pt,height=155pt]{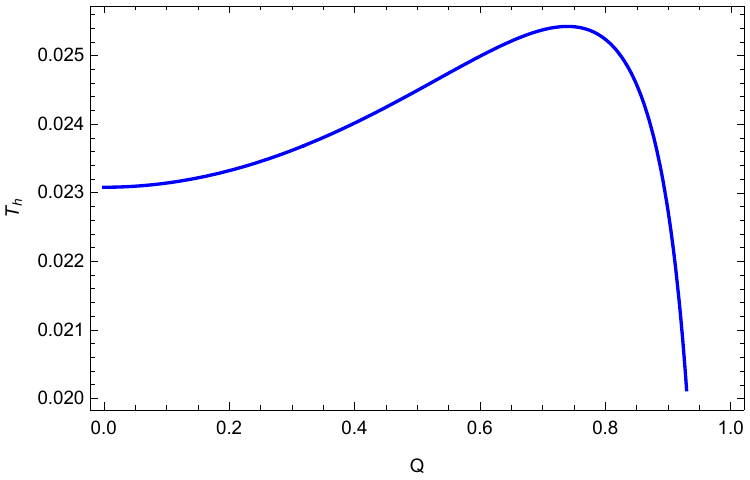}
\label {tempQ}
}
\caption{Hawking temperature for $\Lambda=0.05$, $M=1$; (a) varying $L$ with $Q=0.3$; (b) varying $Q$ with $L=0.2$. }
\label{totaltemp}
\end{figure*} 

\begin{figure*}[!htbp]
\centering
\subfloat[]
{\includegraphics[width=175pt,height=155pt]{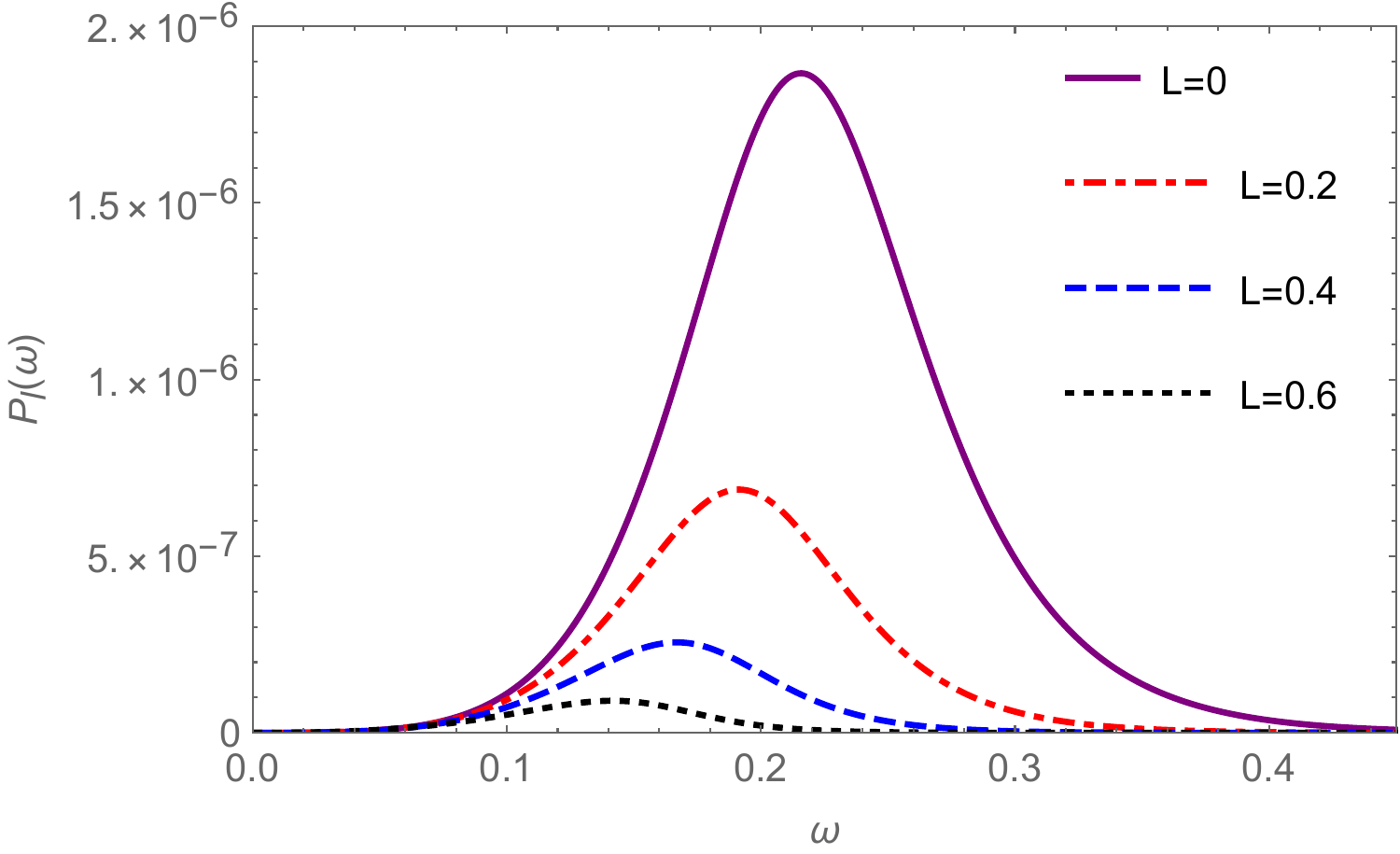}
\label {psL}
}
\hfill
\subfloat[]
{\includegraphics[width=175pt,height=155pt]{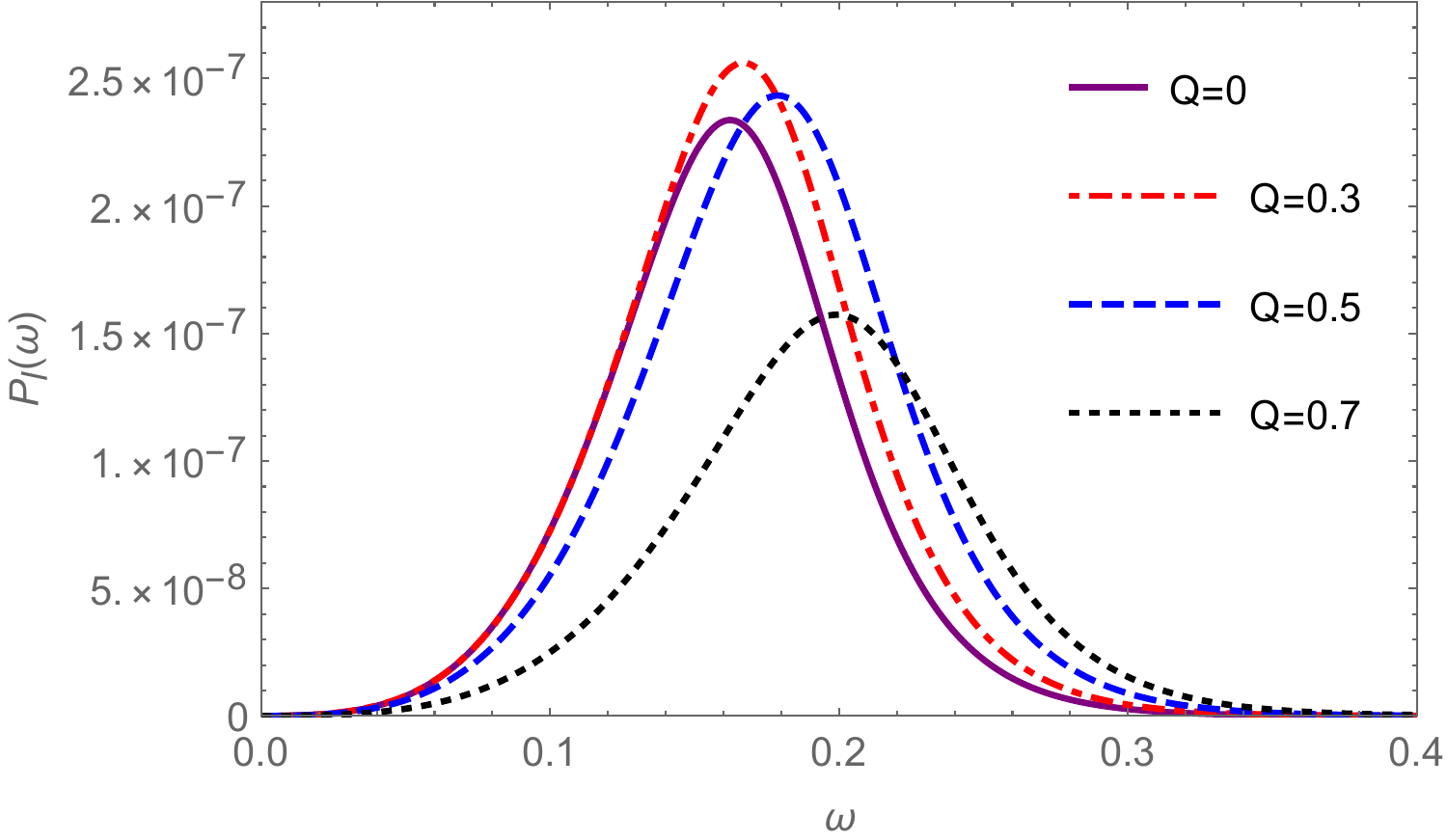}
\label {psQ}
}
\caption{Power spectrum versus $\omega$ with fixed $\Lambda=0.05$, $q=0.1$,   $M=1$.  (a) Variation in $L$ for $Q=0.3$. (b) Variation in $Q$ for $L=0.4$.}
\label{totalps}
\end{figure*}

The dependence of the Hawking temperature $T$ on the parameters $L$ and $Q$ are depicted in Figs. \ref{tempL} and \ref{tempQ} respectively. Increasing $L$ decreases the Hawking temperature monotonically but increasing $Q$ increases the Hawking temperature whereas the Hawking temperature increases initially and reaches upto maximum height and then falls suddenly with increasing $Q$. 
The height of power spectrum suppresses with the increase of $L$ and its peak shifts toward the low frequencies as shown in Fig. \ref{psL}. It is noted from Fig. \ref{psQ} that the height of the power spectrum tends to increases initially with $Q$ and then decreases while the position of peak consistently moves toward higher frequencies.
The numerically calculated values of $\omega_{max}$, $P_{tot}$ and $\eta$ are shown in Tables \ref{PowspectraL} and \ref{PowspectraQ} for massless Dirac field for varying $L$ and $Q$. It shows the effect of $L$ and $Q$ on the sparsity of Hawking radiation of RNdS-like BH. Table \ref{PowspectraL} shows the Hawking radiation become more sparse with the increase of $L$ i.e. the duration between   emissions of radiation quanta increases with increasing $L$. However, we observe from Table \ref{PowspectraQ} that the sparsity initially decreases and then increases with increasing $Q$.

\begin{table}[htp]
\caption{Numerical values of $\omega_{max},  P_{tot}$ and $\eta$ for Dirac perturbation for varying $L$ with fixed    $\Lambda=0.05$, $Q=0.3$, $\ell=1$, $q=0.1$,  and $M=1$.}
\begin{tabular}{p{1cm} p{1.5cm} p{2.5cm}  p{1.5cm}}
\toprule
L &  $\omega_{max} $ &  $P_{tot}$ &  $\eta$ \\ \midrule
0    & 0.216  & $2.27507 \times 10^{-6}$  &  3263.88    \\
0.2  &  0.191 & $1.04379 \times10^{-6}$  &   5562.53    \\
0.4  & 0.167  &  $3.97207\times10^{-7}$ &  11174.7     \\
0.6  & 0.142  &  $ 1.76695\times10^{-7}$ &   18162.3     \\ \bottomrule
\end{tabular}\label{PowspectraL}
\end{table}

\begin{table}[htp]
\caption{Numerical values of $\omega_{max},  P_{tot}$ and $\eta$ for Dirac perturbation for varying $Q$ with fixed $\ell=1$,  $L=0.4$, $q=0.1$, $\Lambda=0.05$,   and $M=1$.}
\begin{tabular}{p{1cm} p{1.5cm} p{2.5cm}  p{1.5cm}}
\toprule
$Q$ &  $\omega_{max} $ &  $P_{tot}$ &  $\eta$ \\ \midrule
0   		 &  		0.162 &  $3.705506\times10^{-7}$		 &    11272.04  \\
0.3 		&  	0.167	 &  	$3.9720719\times10^{-7}$	 &   11174.7   \\
0.5  	&  	0.179	 &  	$5.170827\times10^{-7}$	 &   9862.02   \\
0.7  	&  	0.199	 & $4.939108\times10^{-7}$ 		 &    12760.79       \\ \bottomrule
\end{tabular}\label{PowspectraQ}
\end{table}

%
%
%
%
%

\section{Conclusion}
In this paper, we investigate the strong gravitational lensing of RN-like BH in bumblebee gravity and also derive the radius of photon sphere, deflection angle of light, lens observables which include the position of relativistic images, the angular separation, the relative magnification and time delay between first and second relativistic images with varying $L$ and $Q$. We obtain the lensing coefficients $\bar{a}$ and $\bar{b}$ and investigate the effects of $L$ and $Q$ on the light's deflection angle. It is noted that in the strong field limit, the deflection angle of light always diverges logarithmically when the impact parameter tends to critical impact parameter, corresponding to photon radius and the deflection angle decreases with increasing $Q$. However, the deflection angle decreases initially with the increase of $L$ and small value of $b$ but increases with the higher values of both $b$ and $L$. The observables $\theta_\infty, s, r_{mag}$, $\theta_{n}^E$ and $\Delta T^s_{2,1}$ are discussed both numerically and graphically with varying $L$ and $Q$. It is found that $\theta_\infty, r_{mag}$, $\theta_{n}^E$ and $\Delta T^s_{2,1}$ weakly decrease for the large values of $Q/M$ and $L$ but $s, \bar{a}$ and  $\bar{b}$ increase with increasing $Q/M$ and $L$. Further, using data from various SMBHs like M87$^*$, Sgr A$^*$, NGC  1332, NGC 4649 etc., we perform  a comparative analysis of strong lensing observables 
for RN-like BH and RNBH. \\
 Based on the 1$\sigma$ EHT constraints on the angular shadow diameter of M87* and Sgr A* for different charge-to-mass ratios $Q/M$, we  obtain the corresponding bounds on the parameter $L$. It is shown that the charged black hole solution in bumblebee gravity is consistent with observations within a limited parameter range. Galaxies with high-mass SMBHs (e.g., M87*, NGC 4649) exhibit the largest absolute differences, making them ideal for testing these effects.\\
The Dirac field perturbation of RNdS-like BH in NP formalism is studied and the effective potential which depends on different BH parameters is also derived. The effective potential with radial coordinate $r$ is plotted for different values of $L$ and $Q$ in Figs. \ref{VmasslessL} and \ref{VmasslessQ}. It seems that the effective potential is strictly positive, exhibiting a single maximum between the event horizon and the cosmological horizon. Comparative analysis reveals the RNdS BH has the highest potential peak, exceeding both the RNdS-like and SdS-like BHs. Furthermore, increasing $L$ leads to a reduction in the peak height, while increasing $Q$ causes it to rise. \\
The QNMs of Dirac perturbation are calculated numerically by using 6th-order WKB  and AIM methods. For both methods, higher values of $L$ lower the decay rate and the oscillation frequency but it has the opposite effect for the increase of $Q$. We also investigate the impact of $L$ and $Q$ on the gravitational-wave detectable mass ranges by LIGO and LISA. Increasing $L$ shifts the detectable mass ranges toward lower values, while higher value of $Q$ extends these ranges to more massive systems. Further, the behaviour of GF for different values of $L$ and $Q$ are investigated by using the effective potential of RNdS-like BH. The result shows that the probability of wave transmission of RNdS-like BH is higher than RNdS BH. It is noted that the transmission of wave increases with the increase of $L$ but decreases with the increase of $Q$ as shown in Figs. \ref{transmissionL} and \ref{transmissionQ}. Additionally with increasing $L$, the amount of the incoming wave absorbed by the BH is higher but it becomes opposite for the increase of $Q$ as shown in Figs. \ref{abspartial} and \ref{totalabs} respectively. The Hawking spectrum and its sparsity are also investigated with varying $L$ and $Q$. It is noted that the power spectrum’s peak diminishes and shifts to lower frequencies when $L$ increases but with the increase of $Q$, its spectra increases initially and then decreases, with the peak's position always moves toward the right.

\end{document}